\documentstyle[11pt]{article}
\normalbaselines
\begin{document}
\bibliographystyle{unsrt}

\newtheorem{theorem}{Theorem}
\newtheorem{lemma}{Lemma}
\newtheorem{proposition}{Proposition}

\def\bea*{\begin{eqnarray*}}
\def\eea*{\end{eqnarray*}}
\def\ba{\begin{array}}
\def\ea{\end{array}}
% --------------------------------------------------------------
% If you want to see the names of the equations and references
% set below \count1=0 otherwise \count1=1
% --------------------------------------------------------------
\count1=1
% --------------------------------------------------------------
\def\be{\ifnum \count1=0 $$ \else \begin{equation}\fi}
\def\ee{\ifnum\count1=0 $$ \else \end{equation}\fi}
\def\ele(#1){\ifnum\count1=0 \eqno({\bf #1}) $$ \else \label{#1}\end{equation}\fi}
\def\req(#1){\ifnum\count1=0 {\bf #1}\else \ref{#1}\fi}
\def\bea(#1){\ifnum \count1=0   $$ \begin{array}{#1}
\else \begin{equation} \begin{array}{#1} \fi}
\def\eea{\ifnum \count1=0 \end{array} $$
\else  \end{array}\end{equation}\fi}
\def\elea(#1){\ifnum \count1=0 \end{array}\label{#1}\eqno({\bf #1}) $$
\else\end{array}\label{#1}\end{equation}\fi}
\def\cit(#1){
\ifnum\count1=0 {\bf #1} \cite{#1} \else 
\cite{#1}\fi}
\def\bibit(#1){\ifnum\count1=0 \bibitem{#1} [#1    ] \else \bibitem{#1}\fi}
\def\ds{\displaystyle}
\def\hb{\hfill\break}
\def\comment#1{\hb {***** {\em #1} *****}\hb }

\newcommand{\TZ}{\hbox{\bf T}}
\newcommand{\MZ}{\hbox{\bf M}}
\newcommand{\ZZ}{\hbox{\bf Z}}
\newcommand{\NZ}{\hbox{\bf N}}
\newcommand{\RZ}{\hbox{\bf R}}
\newcommand{\CZ}{\,\hbox{\bf C}}
\newcommand{\PZ}{\hbox{\bf P}}
\newcommand{\QZ}{\hbox{\rm eight}}
\newcommand{\HZ}{\hbox{\bf H}}
\newcommand{\EZ}{\hbox{\bf E}}
\newcommand{\GZ}{\,\hbox{\bf G}}

\font\germ=eufm10
\def\goth#1{\hbox{\germ #1}}
\vbox{\vspace{38mm}}

\begin{center}
{\LARGE \bf The Onsager Algebra Symmetry of $\tau^{(j)}$-matrices in the Superintegrable Chiral Potts Model }\\[10 mm] 
Shi-shyr Roan \\
{\it Institute of Mathematics \\
Academia Sinica \\  Taipei , Taiwan \\
(email: maroan@gate.sinica.edu.tw ) } \\[25mm]
\end{center}

\begin{abstract}
 We demonstrate that the $\tau^{(j)}$-matrices in the superintegrable chiral
Potts model possess the Onsager algebra symmetry for their degenerate
eigenvalues. The Fabricius-McCoy comparison of functional relations of the 
eight-vertex model for roots of unity and the superintegrable chiral Potts model
has been carefully analyzed by identifying equivalent terms in the
corresponding equations, by which we extract the conjectured relation of
$Q$-operators and all fusion matrices in the eight-vertex model corresponding 
to the $T\hat{T}$-relation in the chiral Potts model.
\end{abstract}
\par \vspace{5mm} \noindent
{\it 1999 PACS}:  05.50.+q, 02.20.tTw, 75.10Jm \par \noindent
{\it 2000 MSC}: 14H70, 14Q05, 82B23  \par \noindent
{\it Key words}: Chiral Potts model, Eight-vertex model, Onsager Algebra. \\[10 mm]

\section*{Introduction}
The aim of this paper is to show the Onsager algebra symmetry of $\tau^{(j)}$-matrices of the chiral Potts model (CMP) \cite{BBP} in the superintegrable case. Inspired by the discussion in \cite{FM04} on the $TQ$-relation of eight-vertex model, we observed in \cite{R04} that the degeneracy of  $\tau^{(2)}_p$-eigenvalues occurs at the superintegrable point $p$. Indeed in \cite{FM04},  Fabricius and McCoy introduced the fusion matrices of eight-vertex model for the "root of unity" 
\be
\widetilde{\eta} = \frac{ m_1 K}{N} , \  \  \ ( {\rm gcd}(m_1, N)=1 ) , 
\ele(8eta)
and proposed the conjecture on functional relations related to the $Q$-matrices, of which the structures strikingly resemble the set of functional equations revealed in the study of  $N$-state CPM  in \cite{BBP}. In this paper, we make a detailed analysis on the comparison of these two theories (superintegrable CPM and eight-vertex model for roots of unity) through the equivalence of operators and quantities appeared in functional equations in an "identical" manner (for the correspondence, see table (\req(8CP)) of this article). The effort enables us to further explore one more conjecture on the relationship between $Q$-operators and the whole set of fusion matrices, the equation (\req(8TT)) of this paper, corresponding to the $T\widehat{T}$-relation in CPM which has played the fundamental role in derivation of functional equations in \cite{BBP}.  
In the special case where the elliptic nome vanishes, the eight-vertex model 
has the six-vertex limit for the root of unity $\tilde{\eta}= \frac{m_1 \pi}{2N}$, in which case there is the $sl_2$-loop algebra symmetry \cite{DFM, FM00, FM001, FM01}. Hence the degeneracy of the six-vertex transfer matrix is mainly contributed from finite-dimensional representations of $sl_2$-loop algebra, which are determined by evaluation parameters as the roots of the Drinfeld polynomial. However the essence of Drinfeld polynomial is encoded in the $Q$-operator, by the analysis made in \cite{FM02} on the eight-vertex $Q$-matrix in connection with the function in \cite{De} on degeneracy of the transfer matrix. The analogy between eight-vertex model and CPM, though only loosely related in the present stage, still enables us to suggest that the degeneracy of $\tau^{(2)}$-operator in the superintegrable CPM, previously indicated in \cite{R04}, should further posses a certain symmetry structure similar to the $sl_2$-loop algebra symmetry in six-vertex model in the root of unity case. In this paper we show that it is indeed the case. We identify the full symmetry algebra for $\tau^{(2)}$, hence all $\tau^{(j)}$, matrices to be the Onsager algebra, a renowned Lie algebra appeared in the seminal paper of Onsager on the free energy solution of the two-dimensional Ising model \cite{O}, (for the Onsager algebra, see \cite{D90, D91, DR, R91} and references therein). In the study of (finite-dimensional) representations of Onsager algebra, a proper and useful realization of Onsager algebra is to identify it with the Lie-subalgebra of $sl_2$-loop algebra fixed by a standard involution, by which the $sl_2$-loop algebra representations (with certain constraint on evaluation parameters) give rise to all the irreducible representations of Onsager algebra  \cite{D90, R91}. From the Onsager algebra structure hidden in the Hamiltonion derived from superintegrable CPM, we identify its role in connection to the symmetry of $\tau^{(2)}$-matrices. Furthermore, through the functional relations and eigenvalue spectrum of the transfer matrix in the superintegrable CPM, Onsager algebra indeed describes the full symmetry for the degeneracy of $\tau^{(j)}$-matrices, where only spin-$\frac{1}{2}$ representations occur in the related $sl_2$-loop algebra representations. Hence, the Onsager algebra symmetry of $\tau^{(2)}$-matrices observed in this article, in contrast with the  $sl_2$-loop algebra symmetry of six-vertex model, further enhances the common features shared by both  superintegrable CPM  and eight-vertex model for the roots of unity cases.

This paper is organized as follows. In Sec. 1, we briefly review the main features of the $N$-state CPM (e.g., \cite{AMPT, BPA} and references therein), summarize the set of functional equations of the model in \cite{BBP}, then  describe the specific form of $\tau^{(2)}$-matrix in \cite{R04} and formulae in the superintegrable case considered in this paper. In Sec. 2, we begin with the formalism and conjectured functional relations in eight-vertex model of roots of unity in \cite{FM04}, then make comparison of all entities appeared in functional relations of the superintegrable CPM and eight-vertex model in a precise and one-to-one corresponding manner, by which a conjectured functional equation in eight-vertex model corresponding to $T\widehat{T}$-relation in CPM naturally ascends. The efforts encourage us to propose a speculation about functional relations in a hypothetical solvable lattice model with the similar structure as in the superintegrable CPM. Based on the common features appeared in superintegrable CPM and eight-vertex model for roots of unity, we outline
the procedure and "ingredients" for a such speculated general scheme. The next two sections consist of main results of this paper on the symmetry algebra of superintegrable CPM. In Sec. 3 we start with some basic facts in Onsager algebra and its relation with superintegrable CPM and the quantum $\ZZ_N$-spin chain Hamiltonian in \cite{GR, HKN}. Using the explicit forms of monodromy matrix in the definition of $\tau^{(2)}$ and the Onsager algebra generators in the $\ZZ_N$-spin chain Hamiltonian, we show the Onsager algebra symmetry of $\tau^{(j)}$-matrices in superintegrable CPM. 
While in Sec. 4 we discuss the Bethe equation of $\tau^{(2)}$-matrix and the explicit results on $\tau^{(j)}$-eigenvalues to which the Onsager algebra symmetry of $\tau^{(j)}$-matrices can be understood from the Bethe Ansatz approach based on eigenvalue spectrum of the  superintegrable CPM transfer matrix. Using the explicit form of eigenvalues of CPM transfer matrix, hence the energy expression of the $\ZZ_N$-spin chain Hamiltonian, one can determine that only spin-$\frac{1}{2}$ representations occur in the Onsager algebra symmetry of $\tau^{(j)}$-matrices with the explicit $\tau^{(j)}$-eigenvalue expression. The mathematical nature of ${\cal P}$-polynomial associated to the Bethe equation of the theory has been rigorously investigated so that the full symmetry structure  for the degeneracy of $\tau^{(j)}$-matrices is revealed in Onsager algebra representations.
We close in Sec. 5 with some concluding remarks.

{\bf Notations.}
To present our work, we prepare some notations. In this
paper, $\ZZ, \RZ, \CZ$ will denote 
the ring of integers, real, complex numbers
respectively, $\ZZ_N=
\ZZ/N\ZZ$,  and ${\rm i} = \sqrt{-1}$. For $N \geq 2$, we fix the $N^{\rm th}$ root of unity, $$\omega= e^{\frac{2 \pi i}{N}} ,
$$
and $\CZ^N $ is the vector space consisting of all $N$-cyclic vectors with the basis $\{ | n \rangle \}_{ n \in \ZZ_N}$. Define  $X, Z$  the 
operators of $\CZ^N $ by the relations, $X |n \rangle = | n +1 \rangle$, $ Z |n \rangle = \omega^n |n \rangle $ for $n \in \ZZ_N$, then the Weyl  
relation and $N^{\rm th}$-power identity property  hold: $XZ= \omega^{-1}ZX$, $X^N=Z^N=1$. 
For a positive integers $n$, we denote
by $\stackrel{n}{\otimes} \CZ^N$ the tensor
product of $n$-copies of the vector space $\CZ^N$.

\section{The $N$-state Chiral Potts Model}
We start with the summary of functional equations of chiral Potts model in \cite{BBP}, (for a short version of explanations, see \cite{R04}), then conduct the discussion in the superintegrable case for later consideration. This will also serve to establish the notation.

In the study of two-dimensional $N$-state CMP, the
"rapidity"  of the statistical model is described by a four-vector ratios $[a, b, c, d]$ in the projective 3-space $ \PZ^3$ satisfying the following equivalent sets of equations,
\begin{eqnarray}
{\goth W}  : \, \  \left\{ \begin{array}{l}
ka^N + k'c^N = d^N , \\
kb^N + k'd^N = c^N . \end{array}  \right.   &
\Longleftrightarrow  &
\left\{ \begin{array}{l}
a^N + k'b^N = k d^N  , \\
k'a^N + b^N =
k c^N , \end{array}  \right.
\label{rapidC}
\end{eqnarray}
where $k, k'$ are parameters with $k^2 + k'^2 = 1$, and $k' \neq \pm 1, 0$. The above relations  define ${\goth W}$ as an algebraic curve of genus $N^3-2N^2+1$. 
We shall confine our discussion of CPM only on the full homogeneous lattice by taking $p=p'$ in \cite{BBP}. Hereafter, elements in $\PZ^3$ will be denoted by $p, q, r, \cdots$ etc., and we shall use the variables $x, y, \mu , t, \lambda$ to denote the following component-ratios for an element $[a, b, c, d] \in \PZ^3$,
$$
x := \frac{a}{d} , \ \ \  y :=  \frac{b}{c} , \ \ \ \mu : =   \frac{d}{c} , \ \ \ \ t : = \frac{a b}{c d } \ (= x y ) \ , \ \ \ \lambda := \frac{d^N}{c^N} \ ( = \mu^N ) \ .
$$
Associated to an element $p$ in $\PZ^3$, the coordinates and variables will be written by $a_p, b_p, x_p, t_p \cdots $ so on whenever if it will be necessary to specify the element $p$. The variables, $x, y, \mu$, form a system of affine coordinates of $\PZ^3$, by which an equivalent form of the defining equation for the curve ${\goth W}$ in (\ref{rapidC}) is 
\be
k x^N  = 1 -  k'\mu^{-N}, \ \ \  k y^N  = 1 -  k'\mu^N \ , \ \ (x , y , \mu ) \in \CZ^3 \ .
\ele(xymu)
The rapidity curve ${\goth W}$ has a large symmetry group (with the order $4N^3$ for $N \geq 3$ \cite{AP, R04}), in which the following automorphisms will be needed in later discussions,
\bea(ll)
U :  [a,b,c,d] \mapsto [\omega a,b,c,d], & \bigg(  (x, y, \mu) \mapsto ( \omega x, y, \mu) \bigg) , \\
C :  [a,b,c,d] \mapsto [b,a, d, c ], & \bigg((x, y, \mu) \mapsto ( y, x , \mu^{-1} ) \bigg)  .
\elea(UC)
Note that $U^N = C^2 = 1$.
By eliminating the variable $\mu^N$ in (\req(xymu)) , ${\goth W}$ becomes a $N$-fold unramified cover of the genus $(N-1)^2$ curve,
\be
x^N + y^N = k ( 1 + x^N y^N )   \ .
\ele(xy) 
By (\req(xymu)), the variables, $(t, \lambda) =(t_p, \lambda_p)$ for $p \in {\goth W}$, define the following hyperelliptic curve of genus $N-1$,
\be
W_{k'} : \ \ t^N = \frac{(1- k' \lambda )( 1 - k' \lambda^{-1}) }{k^2 } \ ,
\ele(Wk')
which is a $N$-ramified quotient of the curve (\req(xy)). The variables $t, \lambda$ in the above curve have played a major role of rapidities for solutions of various problems in CPM,  see e.g. \cite{B90, B05, MR} and references therein.

The Boltzmann weights
$W_{p,q},\overline{W}_{p,q}$ of the $N$-state CPM are defined by coordinates of $p, q \in {\goth W}$ with the expressions: 
$$
\begin{array}{ll}
\frac{W_{p,q}(n)}{W_{p,q}(0)}  = \prod_{j=1}^n
\frac{d_pb_q-a_pc_q\omega^j}{b_pd_q-c_pa_q\omega^j}  & \bigg( = (\frac{\mu_p}{\mu_q})^n \prod_{j=1}^n
\frac{y_q-\omega^j x_p}{y_p- \omega^j x_q }  \bigg), \\
\frac{\overline{W}_{p,q}(n)}{\overline{W}_{p,q}(0)} 
 = \prod_{j=1}^n
\frac{\omega a_pd_q-
d_pa_q\omega^j}{ c_pb_q- b_pc_q \omega^j}  & \bigg( = ( \mu_p\mu_q)^n \prod_{j=1}^n \frac{\omega x_p - \omega^j x_q }{ y_q- \omega^j y_p }  \bigg) \ .
\end{array}
$$
By the rapidity constraint (\ref{rapidC}), the above Boltzmann weights have the $N$-periodicity property for $n$. Without loss of generality, we may assume $W_{p, q}(0) = \overline{W}_{p,q}(0) = 1$.
On a lattice of the horizontal size $L$ with 
periodic boundary condition, the combined weights of
intersection between two consecutive rows give rise
to an operator of $\stackrel{L}{\otimes} \CZ^N$, which defines the transfer matrix of the $N$-state CPM:
\begin{eqnarray}
T_p (q)_{\sigma, \sigma'} = \prod_{l=1}^L
\overline{W}_{p,q}(\sigma_l - \sigma_l')
W_{p,q}(\sigma_l - \sigma_{l+1}') \ , \ \ \ p, q \in {\goth W} \ ,
\label{Tpq}
\end{eqnarray}
where $\sigma= ( \sigma_1 , \ldots , \sigma_L) $ ,
$\sigma'= ( \sigma_1' , \ldots , \sigma_L') $ with $\sigma_l, \sigma_l' \in
\ZZ_N$. By the
star-triangle relation of Boltzmann weights:
$$
\sum_{n=0}^{N-1} \overline{W}_{qr}(\sigma' - n) W_{pr}(\sigma - n) \overline{W}_{pq}(n - \sigma'') = R_{pqr} W_{pq}(\sigma - \sigma' )  \overline{W}_{pr}(\sigma' - \sigma'') W_{qr}(\sigma -\sigma'') \   
$$
with $R_{pqr}= \frac{f_{pq}f_{qr}}{f_{pr}}$ and $f_{pq}= \bigg( \frac{{\rm det}_N( \overline{W}_{pq}(i-j))}{\prod_{n=0}^{N-1} W_{pq}(n)}\bigg)^{\frac{1}{N}}$, one has the commutativity of  
transfer matrices for a fixed $p \in {\goth W} $:
$$
[ T_p (q) , \ \ T_p (q') ] = 0 \  \ , \ \ \ q , q' \in {\goth W} \ . 
$$
By (\req(xymu)), $T_p (q)$ for a fixed $p$ depends only on the values of $(x_q, y_q)$, parametrized by the curve (\req(xy)). Hence we shall also write $T_p(q)$ by $T_p (x_q, y_q)$ whenever this is convenient. The $\ZZ_N$-spin operators $X, Z$ of $\CZ^N$ give rise to the system of Weyl operators of $\stackrel{L}{\otimes} \CZ^N$:
$$
X_j = 1 \otimes \cdots \otimes X^{jth} \otimes 1 \otimes \cdots \otimes 1 , \ \ \ Z_j = 1 \otimes \cdots \otimes Z^{jth} \otimes 1 \otimes \cdots \otimes 1, \ \ \ \ (1 \leq j \leq L) 
$$
with $Z_i X_j = \omega^{\delta_{i, j}} X_j Z_i$, $[Z_i, Z_j]= [X_i, X_j]=0$ and $Z_j^N= X_j^N=1$.  The spin-shift operator of $\stackrel{L}{\otimes} \CZ^N$ will be again denoted by 
$X =\prod_{j=1}^L X_j$ if no confusion could arise. 
Then $T_p(q)$ commutes with $X$ and the spatial translation operator $S_R$, which takes the $j$th column to $(j+1)$th one for $1 \leq j \leq L$ with the 
identification $L+1 = 1$. The eigenvalues of $X$ and $ S_R$, denoted by $e^{{\rm i}Q}, e^{{\rm i}P}$ for $Q \in \ZZ_N, P \in \ZZ_L$, are quantum numbers, called the $\ZZ_N$-charge and total momentum respectively. We denote
$$
\widehat{T}_p(q) = T_p(q) S_R \ . 
$$

In the study of CPM as a descendent of the six-vertex model, Bazhanov and Stroganov discovered a five-parameter family of
solutions of Yang-Baxter equation associated to the six-vertex R-matrix with entries in terms of operators $X, Z$  acting on "quantum space" $\CZ^N$ \cite{BazS, GPS}. In particular, there is a Yang-Baxter solution $G(t)$ associated to an element $p=[a, b, c, d] \in \PZ^3$ with the form \cite{R04}:
\bea(l)
 b^2 G_p (t)  = \left( \begin{array}{cc}
       b^2 - t d^2 X & ( bc - \omega a d X )  Z   \\
       -t (b c - a d X ) Z^{-1} & - t c^2  + \omega  a^2  X  
\end{array} \right)  \ , \ \ \
t \in \CZ \ ,
\elea(G)
which satisfy the Yang-Baxter relation:
\be
R(t/t') (G_p (t) \bigotimes_{aux}1) ( 1
\bigotimes_{aux} G_p(t')) = (1
\bigotimes_{aux} G_p(t'))(G_p (t)
\bigotimes_{aux} 1) R(t/t') \ ,
\ele(YBe)
where $R(t)$ is the following matrix of 2-tensor of 
the "auxiliary  space" $\CZ^2$,
$$
R(t ) = \left( \begin{array}{cccc}
        t \omega - 1  & 0 & 0 & 0 \\
        0 &t-1 & \omega  - 1 &  0 \\ 
        0 & t(\omega  - 1) &\omega( t-1) & 0 \\
     0 & 0 &0 & t \omega - 1    
\end{array} \right) \ .
$$
By the auxiliary-space matrix-product and quantum-space tensor-product for a finite size $L$, the operator 
\be
\bigotimes_{\ell =1}^L G_{p, \ell} (t)  = G_{p, 1} (t) \otimes \cdots \otimes G_{p, L} (t) \ , \ \ \ \ 
G_{p, \ell} (t) :=  G_p (t) \ {\rm at \ site \ } \ell \ ,
\ele(Gj) 
again satisfies the Yang-Baxter relation (\req(YBe)), hence the traces, ${\rm tr}_{aux} \bigotimes_{\ell=1}^L G_{p, \ell} (t)$ ~ $ (t \in \CZ)$, form a
family  of commuting operators of  $\stackrel{L}{\otimes} \CZ^N$. The $\tau^{(2)}_p$-operator is defined by
\be
\tau^{(2)}_p (t) = {\rm tr}_{aux} ( \bigotimes_{\ell=1}^L G_{p, \ell} ( \omega t)) \ \ \ \ {\rm for} \ \  t \in \CZ . 
\ele(tau2)

When $p \in {\goth W}$, the transfer matrix $T_p (q)$ of CPM can be derived from $\tau^{(2)}_p (t_q)$ as the auxiliary $"Q"$-operator in a similar way as in the TQ-relation discussion of eight-vertex model in \cite{Bax}.  One has the following $\tau^{(2)}T$-relation ((4.20) in \cite{BBP}, or (14) in \cite{B02})\footnote{$\tau^{(2)}_p ( q ), T_p (q)$ in this paper are the operators $\tau^{(2)}_{k=0, q}$, $T_q$ in \cite{BBP} respectively.}: 
\be
\tau^{(2)}_p(t_q) T_p(\omega x_q, y_q) = \varphi_p(q) T_p(x_q, y_q) + \overline{\varphi}_p(Uq) X T_p(\omega^2x_q, y_q) \ , 
\ele(tauT)
where $
\varphi_p(q):= (\frac{(y_p -\omega x_q)(t_p -t_q)}{y_p^2(x_p - x_q)})^L$, $ \overline{\varphi}_p(q) := (\frac{\omega \mu_p^2(x_p -  x_q)(t_p - t_q)}{y_p^2(y_p - \omega x_q)})^L $, and 
$U$ is the automorphism of ${\goth W}$ in (\req(UC)).
An equivalent form of the $\tau^{(2)}T$-relation (\req(tauT)) is the following expression of $\tau^{(2)}_p$ in terms of $T_p$:
\be
\tau^{(2)}_p(t_q) = \bigg( \varphi_p(q) T_p(x_q, y_q) + \overline{\varphi}_p(Uq) X T_p(\omega^2x_q, y_q) \bigg) T_p(\omega x_q, y_q)^{-1} \ ,
\ele(tau2T)
by which, the commutativity of $T_p(q)$'s ensures 
\be
[ \tau^{(2)}_p(t_q) , T_p( x_{q'}, y_{q'}) ] = 0 \ , \ \ \ \ {\rm for} \ \ p, q, q' \in {\goth W} \ .
\ele(tau2T0)
There are operators $\tau^{(j)}(t)$ for $0 \leq j \leq N$ with $\tau^{(0)}_p(t):= 0 $, $\tau^{(1)}_p(t):= I$, which are  related to transfer matrices $T_p(q)$ by the following $T\widehat{T}$-relations ((3.46) for $(l,k)=(j, 0)$ in \cite{BBP}, or (13) in \cite{B02}):
\be
T_p(x_q, y_q) \widehat{T}_p( y_q, \omega^j x_q) =  r_{p, q} h_{j ; p, q} \bigg( \tau^{(j)}_p(t_q) + \frac{z(t_q)z(\omega t_q)\cdots z(\omega^{j-1} t_q)}{\alpha_p (\lambda_q)} X^j \tau_p^{(N-j)} (\omega^j t_q) \bigg) 
\ele(TThat)
for $0 \leq j \leq N$, where $r_{p, q}  = \bigg(\frac{N(x_p - x_q) (y_p - y_q) (t_p^N-t_q^N)}{(x_p^N - x_q^N) (y_p^N - y_q^N)(t_p - t_q) } \bigg)^L$ , $
h_{j ; p, q} =  \bigg( \prod_{m=1}^{j-1} \frac{y_p^2 (x_p - \omega^m x_q)}{(y_p - \omega^m x_q)(t_p - \omega^m t_q) }\bigg)^L $ and 
\bea(lll)
z(t)&= \bigg(\frac{\omega \mu_p^2 (x_p y_p - t )^2 }{y_p^4}\bigg)^L &(= \varphi_p(q) \overline{\varphi}_p(q)) , \\
\alpha_p ( \lambda_q ) &= \bigg(\frac{k'(1-\lambda_p \lambda_q)^2}{ \lambda_q (1-k' \lambda_p)^2 }\bigg)^L &(= \bigg(\frac{(y_p^N-x_q^N)(t_p^N-t_q^N)}{ y_p^{2N}(x_p^N-x_q^N) }\bigg)^L
).
\elea(zalpha) 
In particular, the relation (\req(TThat)) for $j=N$ (or 0) becomes
\be
T_p(x_q, y_q) \widehat{T}_p( y_q,  x_q) =  \bigg( \frac{N y_p^{2N-2}(y_p - y_q)(y_p -  x_q) }{ (y_p^N - y_q^N)(y_p^N - x_q^N ) } \bigg)^L  \tau^{(N)}_p(t_q)  \ .  
\ele(TThatN) 
By (\req(tau2T0)) and (\req(TThat) ), the following "fusion relations" hold for $\tau^{(j)}$'s ( (4.27) of \cite{BBP} ):
\bea(cl)
\tau^{(j)}_p(t) \tau^{(2)}_p(\omega^{j-1} t) = z( \omega^{j-1} t ) X \tau^{(j-1)}_p(t) + \tau^{(j+1)}_p(t)  , & 1 \leq j \leq N \ , \\
\tau^{(N+1)}_p(t):= z(t ) X \tau^{(N-1)}_p( \omega t) + u (t) I \ ,
\elea(Fus)
where $u (t) = \alpha_p ( \lambda) + \alpha_p ( \lambda^{-1} )$, and $z(t)$ in (\req(zalpha)). Therefore the fusion operators $\tau^{(j)}_p$ can be constructed recursively from $\tau^{(2)}_p$ and relations in (\req(Fus)) by setting $\tau^{(0)}_p = 0 $, $\tau^{(1)}_p = I$, and one can express  $\tau^{(j)}_p (t)$ as a "polynomial" of $\tau^{(2)}_p$ of degree $(j-1)$ with coefficients in powers of $X$. Indeed the explicit form of $\tau^{(j)}_p(t)$ for $ 2 \leq j \leq N+1$ in terms of $\tau^{(2)}_p $ and $X$ is as follows \cite{R04}: 
\begin{eqnarray}
\tau^{(j)}_p(t)  = \prod_{s=0}^{j-2} \tau^{(2)}_p(\omega^s t) + \sum_{k=1}^{[\frac{j-1}{2}] } (-X)^k  \sum_{1 \leq i_1 <'i_2 <' \cdots <' i_k \leq j-2}\bigg( \prod_{\ell=1}^k \frac{z(\omega^{i_\ell} t)}{\tau^{(2)}_p(\omega^{i_\ell-1 }t )\tau^{(2)}_p(\omega^{i_\ell }t)} \prod_{s=0}^{j-2} \tau^{(2)}_p(\omega^s t) \bigg) \  \label{tauF}
\end{eqnarray}
where $i_\ell <' i_{\ell+1}$ means $i_\ell + 1 < i_{\ell+1}$.

Using (\req(tau2T)) and (\req(Fus)), one can successively express  $\tau_p^{(j)}$ in terms of $T_p(q)$, then obtain the following $\tau^{(j)}T$-relations for $1 \leq j \leq N+1$, ((4.34) in \cite{BBP}):
\bea(ll)
\tau^{(j)}_p(q) =&  T_p(x_q, y_q) T_p(\omega^j x_q , y_q) \sum_{m=0}^{j-1} \bigg( \varphi_p(q)\varphi_p(Uq) \cdots \varphi_p(U^{m-1}q)  \times \\
&\overline{\varphi}_p(U^{m+1}q) \cdots \overline{\varphi}_p(U^{j-1}q) 
T_p(\omega^m x_q, y_q)^{-1} T_p(\omega^{m+1}x_q, y_q)^{-1} X^{j-m-1} \bigg) .
\elea(taujT)
Indeed,  by $\prod_{j=0}^{N-1} \varphi_p(U^j q) = \alpha_p (\lambda_q)$ and $\prod_{j=0}^{N-1} \overline{\varphi}_p(U^j q) = \alpha_p (\lambda_q^{-1})$, $\tau^{(2)}T$-relation with the fusion relation is equivalent to the $\tau^{(j)}T$-relations : $(\req(tauT))+ (\req(Fus)) \Longleftrightarrow (\req(taujT))$. By (\req(TThatN)) and $(\req(taujT))_{j=N}$, one obtains the functional equation of chiral Potts transfer matrices $T_p$ ((4.40) of \cite{BBP}):
\bea(ll)
 \widehat{T}_p( y_q,  x_q) &=    \sum_{m=0}^{N-1} C_{m; p}(q) T_p( x_q , y_q) T_p(\omega^m x_q, y_q)^{-1} T_p(\omega^{m+1}x_q, y_q)^{-1} X^{-m-1} ,
\elea(TT)
where $C_{m; p}(q) =   \varphi_p(q)\varphi_p(Uq) \cdots \varphi_p(U^{m-1}q) 
\overline{\varphi}_p(U^{m+1}q) \cdots \overline{\varphi}_p(U^{N-1}q) ( \frac{N y_p^{2N-2}(y_p - y_q)(y_p -  x_q) }{ (y_p^N - y_q^N)(y_p^N - x_q^N ) })^L $.
%%$$ \widehat{T}_p( y_q,  x_q) =   
% \sum_{m=0}^{N-1} C_{m; p}^\prime (q) T_p( x_q , y_q)  T_p(\omega^{-m}x_q, y_q)^{-1} %T_p(\omega^{-m-1} x_q, y_q)^{-1} X^m $$
%where $C_{m; p}^\prime = C_{N-1-m; p}(q) = \overline{\varphi}_p(U^{-1}q) \cdots %\overline{\varphi}_p(U^{-m}q) \varphi_p(U^{-m-2}q) \cdots \varphi_p(U^{-N}q)   ( \frac{N %y_p^{2N-2}(y_p - y_q)(y_p -  x_q) }{ (y_p^N - y_q^N)(y_p^N - x_q^N ) })^L $ 
Note that substitutions of (\req(taujT)) in (\req(TThat)) all give the same relation (\req(TT)). Hence under the scheme of $\tau^{(2)}T$-relation (\req(tauT)), the $T_p$-functional equation (\req(TT)) is equivalent to $T\widehat{T}$-relations (\req(TThat)). The functional equations (\req(Fus)) and (\req(TThat)) have provided an effective method to study the eigenvalue problem of CPM by reducing the relations in forms with the variables $(t, \lambda)$ in (\req(Wk')) using the following formula ((2.40) in \cite{BBP}):
\be
T_p ( \omega x_q , \omega^{-1} y_q ) = \bigg( \frac{(y_p - \omega x_q)(y_p - \omega^{-1} y_q)}{\mu_p^2 (\omega x_p - y_q) (x_p - x_q )} \bigg)^L X^{-1} T_p (x_q, y_q) \ . 
\ele(TM0)

Hereafter, the chiral Potts model considered in this paper will only be the superintegrable case, i.e., the vertical rapidity $p$ is   given by
\be
p : \ x_p = y_p = \eta^{\frac{1}{2}} \ , \ \ \mu_p = 1 \ , \ \ \ \ {\rm where} \ \ \eta = ( \frac{1-k'}{1+k'})^{\frac{1}{N}}.
\ele(Sip)
We shall omit the subscript $p$ of the operators $\tau^{(j)}_p$, $T_p(x_q, y_q)$, $ \widehat{T}_p ( x_q , q_q )$ for $p$ in (\req(Sip)), e.g. $\tau^{(j)}$ means $\tau^{(j)}_p$, etc. 
Then monodromy matrix in (\req(G)) becomes
\bea(l)
 G (t)  = \left( \begin{array}{cc}
       G_0^0 (t) & G_0^1 (t)  \\
       G_1^0 (t) & G_1^1 (t)  
\end{array} \right) = \left( \begin{array}{cc}
       1 - \eta^{-1} t X & \eta^{\frac{-1}{2}}( 1 - \omega  X )  Z   \\
       -\eta^{\frac{-1}{2}} t (1 - X ) Z^{-1} & - \eta^{-1}t   + \omega   X  
\end{array} \right)  \ ,
\elea(Gsi)
and $\tau^{(2)}$-matrix (\req(tau2)) is expressed by
\be
\tau^{(2)} (t ) = {\rm tr}_{aux} ( \bigotimes_{\ell=1}^L G_\ell) ( \omega t) = \bigg( \sum_{\alpha_1, \alpha_2, \ldots, \alpha_L } (G_1)_{\alpha_1}^{\alpha_2}(G_2)_{\alpha_2}^{\alpha_3} \cdots (G_L)_{\alpha_L}^{\alpha_1} \bigg) ( \omega t) 
\ele(tau2Si)
where $\alpha_\ell = 0, 1$ for all $\ell$. The fusion relation (\req(Fus)) becomes 
\bea(cll)
\tau^{(j)}(t) \tau^{(2)}(\omega^{j-1} t) &= (1 - \eta^{-1} \omega^{j-1} t )^{2L} ~  \tau^{(j-1)}(t) ~ \omega^L X ~ + \tau^{(j+1)}(t)  , & 1 \leq j \leq N \ ,  \\
\tau^{(N+1)}(t) & = (1 - \eta^{-1}  t )^{2L} ~ \tau^{(N-1)}( \omega t) \omega^L X  ~ + 2 (1 - \eta^{-N} t^N  )^L  \ . &
\elea(SFus)
%\bea(cl)\tau^{(j)}(t) \tau^{(2)}(\omega^{j-1} t) = {\bf z}( \omega^{j-1} t )  %\tau^{(j-1)}(t) \omega^L X + \tau^{(j+1)}(t)  , & 1 \leq j \leq N \ ,  \\
%\tau^{(N+1)}(t)= {\bf z}(t ) \tau^{(N-1)}( \omega t) \omega^L X  + {\bf u}(t) I \ &
%\elea(SFus) where ${\bf z}(t)= (1 - \eta^{-1} t )^{2L} $, and ${\bf u} (t) = 
% 2 (1 - \eta^{-N} t^N  )^L $. 
By (\ref{tauF}), one has 
\begin{eqnarray}
\tau^{(j)}(t)  = \prod_{s=0}^{j-2} \tau^{(2)}(\omega^s t) + \sum_{k=1}^{[\frac{j-1}{2}] } (-\omega^L X)^k  \sum_{1 \leq i_1 <'i_2 <' \cdots <' i_k \leq j-2}\bigg( \prod_{\ell=1}^k \frac{(1 - \omega^{i_\ell}\eta^{-1} t )^{2L}}{\tau^{(2)}(\omega^{i_\ell-1 }t )\tau^{(2)}(\omega^{i_\ell }t)} \prod_{s=0}^{j-2} \tau^{(2)}(\omega^s t) \bigg) \  \label{SitauF}
\end{eqnarray}
for $2 \leq j \leq N+1$.

Define 
\be
{\sc Q}_{cp} (q) = \frac{T (x_q, y_q)(1- \eta^{\frac{-N}{2}} x_q^N)^L}{ N^L (1 - \eta^{\frac{-1}{2}} x_q )^L} , \ \ \ \ \ \widehat{\sc Q}_{cp} (q) = \frac{\widehat{T} (x_q, y_q)(1- \eta^{\frac{-N}{2}} x_q^N)^L}{ N^L (1 - \eta^{\frac{-1}{2}}x_q )^L} \ .
\ele(Qcp)
Since  $
\frac{T ( \omega x_q , \omega^{-1} y_q )}{(1 - \eta^{\frac{-1}{2}}\omega x_q)^L} = \frac{T (x_q, y_q)}{ (1 - \eta^{\frac{-1}{2}} x_q )^L} (\omega^L X)^{-1}$ by (\req(TM0)), 
 ${\sc Q}_{cp} (q)^N$ and $\widehat{\sc Q}_{cp} (q)^N$ depend only on the value of $(t_q,\lambda_q)$ in the curve (\req(Wk')). Hence, up to a $N$th root of unity, we may write ${\sc Q}_{cp} (q), \widehat{\sc Q}_{cp} (q)$ simply by ${\sc Q}_{cp} (t_q, \lambda_q)$, $\widehat{\sc Q}_{cp} (t_q, \lambda_q)$ as well. 
Then $\tau^{(2)}T$-relation (\req(tauT)) becomes
\be
\tau^{(2)}(t_q) {\sc Q}_{cp} (\omega t_q, \lambda_q) = (1 -\eta^{-1} t_q)^L {\sc Q}_{cp} (t_q, \lambda_q) + (1 - \eta^{-1} \omega t_q)^L  {\sc Q}_{cp} (\omega^2 t_q, \lambda_q) \omega^L X \ ,  
\ele(tauTSi)
and $\tau^{(j)}T$-relations (\req(taujT)) have the form 
\begin{eqnarray}
&\tau^{(j)}(t_q) 
= &  \sum_{m=0}^{j-1} \bigg( \frac{\prod_{ k=0}^{j-1} (1- \omega^k \eta^{-1}t_q)^L}{(1- \omega^m \eta^{-1}t_q)^L} {\sc Q}_{cp}(t_q, \lambda_q)
{\sc Q}_{cp}(\omega^m t_q, \lambda_q)^{-1} \times \nonumber \\
& &
{\sc Q}_{cp}(\omega^j t_q , \lambda_q) {\sc Q}_{cp}(\omega^{m+1}t_q, \lambda_q)^{-1} (\omega^L X)^{j-m-1} \bigg)  .
\label{taujTSi}
\end{eqnarray}
The $T\widehat{T}$-relations, (\req(TThat)) and (\req(TThatN)), become 
\begin{eqnarray}
{\sc Q}_{cp}(t_q, \lambda_q) \widehat{\sc Q}_{cp}( \omega^j t_q, \lambda_q^{-1})  =& \frac{(1- \eta^{-N} t_q^N)^L}{N^L } \bigg( \frac{1}{\prod_{m=0}^{j-1} (1 - \omega^m \eta^{-1} t_q)^L } \tau^{(j)}(t_q)  & \nonumber \\
& + \frac{ (\omega^L X)^j }{\prod_{m=j}^{N-1} (1 - \omega^m \eta^{-1} t_q)^L} \tau^{(N-j)} (\omega^j t_q)  \bigg) \ , & 0 \leq j \leq N , \label{TTS} \\
{\sc Q}_{cp}(t_q, \lambda_q) \widehat{\sc Q}_{cp}( t_q, \lambda_q^{-1} )  =& \frac{1}{N^L}  \tau^{(N)}(t_q) \ . \ \ \ \ \ \ \ \ \ \ \ \ \ \ \ & 
\label{TTNS}
\end{eqnarray}
By (\req(TT)), one has the following functional equation of ${\sc Q}_{cp}(q)$:
\begin{eqnarray}
\widehat{\sc Q}_{cp}( C q ) = \frac{(1- \eta^{-N} t_q^N)^L}{N^L }   \sum_{m=0}^{N-1} \frac{(\omega^L X)^{-1-m} }{(1- \omega^m \eta^{-1} t_q)^L}  {\sc Q}_{cp}(q) {\sc Q}_{cp}(U^m q)^{-1}{\sc Q}_{cp}(U^{m+1}q)^{-1} 
\label{QS}
\end{eqnarray}
where $C, U$ are the transformations in (\req(UC)).
%$$
%\widehat{\sc Q}_{cp}( C q ) =  \bigg(\frac{N(1- \eta^{-N} t_q^N)}{(1 - \eta^{\frac{-N}{2}} 
%x_q^N ) (1 - \eta^{\frac{-N}{2}} y_q^N) }\bigg)^L   \sum_{m=0}^{N-1} 
%\frac{(\omega^L X)^m }{(1- \omega^{-m-1} \eta^{-1} t_q)^L}  {\sc Q}_{cp}(q) 
%{\sc Q}_{cp}(U^{-m}q)^{-1} {\sc Q}_{cp}(U^{-m-1} q)^{-1} . $$

\section{Eight-Vertex Model for the Root of Unity Cases}
In \cite{FM04}, Fabricious and McCoy observed the similar structures between the eight-vertex model for the "root of unity" cases and the superintegrable $N$-state CPM. They derived and proposed a set of functional equations for the eight-vertex model analogous to those of CPM in \cite{BBP} as follows. 

The transfer matrix for the eight-vertex model in the root of unity $\widetilde{\eta}$ in (\req(8eta)) with {\it even} $L$ columns and the periodic boundary condition is\footnote{The $L, N$ in \cite{FM04} are changed to $N, L$ for the parallel notations of the chiral Potts model used in this paper.} 
$$
T^{(2)}(v)_{| \mu, \nu } = {\rm Tr} W_8( \mu_1, \nu_1 )W_8( \mu_2, \nu_2 ) \cdots W_8( \mu_L, \nu_L )  
$$ 
where the Boltzmann weights are parametrized by elliptic theta functions\footnote{Here we use the standard conventions for Jacobi theta functions $H(v), \Theta (v)$ with  elliptic integrals $K, K'$ of nome $q = e^{\frac{- \pi K'}{K}}$: $
H(v) = 2 \sum_{n=1}^\infty (-1)^{n-1} q^{(n-1/2)^2} \sin \frac{(2n-1) \pi v}{2K}$, $ \Theta (v) = 1 + 2 \sum_{n=1}^\infty (-1)^n q^{n^2} \cos \frac{n \pi v}{K }$. },
$$
\begin{array}{llll}
W_8(1, 1)_{|1,1} &= W_8(1, 1)_{|1,1} &= \Theta ( 2 \widetilde{\eta})\Theta (v- \widetilde{\eta}) H (v+ \widetilde{\eta}) &= a(v) , \\
W_8(-1, -1)_{|1,1} &= W_8(1, 1)_{|-1,-1} &= \Theta ( 2 \widetilde{\eta}) H (v- \widetilde{\eta}) \Theta (v+ \widetilde{\eta}) &= b(v) , \\
W_8(-1, 1)_{|1,-1} &= W_8(1, -1)_{|-1,1} &= H ( 2 \widetilde{\eta})\Theta (v- \widetilde{\eta}) \Theta (v+ \widetilde{\eta}) &= c(v) , \\
W_8(1, -1)_{|1,-1} &= W_8(-1, 1)_{|-1,1} &= H ( 2 \widetilde{\eta})H (v- \widetilde{\eta}) H (v+ \widetilde{\eta}) &= d(v) .
\end{array}
$$
In \cite{B72}, Baxter defined the "auxiliary" $Q$-operators, $Q_R(v)$ and its companion $Q_L(v)$, to study the eigenvalue problem of $T^{(2)}(v)$ through the following $TQ$-relations: 
$$
\begin{array}{l}
T^{(2)}(v) Q_R (v) = h^L( v - \widetilde{\eta} ) Q_R ( v + 2 \widetilde{\eta} ) + h^L ( v+ \widetilde{\eta}) Q_R(v- 2 \widetilde{\eta} ) , \\
Q_L (v) T^{(2)}(v)  = h^L( v - \widetilde{\eta} ) Q_L ( v + 2 \widetilde{\eta} ) + h^L ( v+ \widetilde{\eta}) Q_L (v- 2 \widetilde{\eta}) , 
\end{array}
$$
where $h(v) = \Theta (0) \Theta (v) H (v)$. Furthermore, the following relations hold:
$$
\begin{array}{cl}
T(v + N \widetilde{\eta}) = T(v) , & h(v + N \widetilde{\eta}) = h (v)  , \\
 Q_L (v) Q_R (v') = Q_R (v') Q_L (v) , & Q_{R, L} (v + N \widetilde{\eta}) = S^{m_1} Q_{R, L} (v)= Q_{R, L} (v) S^{m_1}
\end{array}
$$
where $S = \prod_{j = 1}^L \sigma_j^z$. Define $Q_{8V}(v) = Q_R (v) Q_R(v_0)^{-1} = Q_L(v_0)^{-1} Q_L(v)$ for a base element $v_0$. Then $\{ Q_{8V}(v) \}_v$ and $\{ T^{(2)}(v') \}_{v'}$ are commutative families of operators such that $[T^{(2)}(v'), Q_{8V}(v)]=0$ and the following $TQ$-relation holds:
$$
T^{(2)}(v) Q_{8V} (v) = h^L( v - \widetilde{\eta} ) Q_{8V} ( v + 2 \widetilde{\eta} ) + h^L ( v+ \widetilde{\eta}) Q_{8V}(v- 2 \widetilde{\eta} ) \ ,
$$
which is equivalent to
\be
T^{(2)}( v + 2 \widetilde{\eta} ) Q_R ( v + 2 \widetilde{\eta}) = h^L ( v + 3 \widetilde{\eta} ) Q_R (v) + h^L( v+ \widetilde{\eta} ) Q_R( v+ 4 \widetilde{\eta}) \ .
\ele(8TQ)
In \cite{FM04}, Fabricius and McCoy defined 
the "fusion matrices" $T^{(j)}(v)$ recursively by setting $T^{(0)}(v)= 0$, $T^{(1)}(v+ 2 \widetilde{\eta})= h^L( v + \widetilde{\eta})$, and the following fusion relations ((3.4) (3.5) (3.6) and (3.15) in \cite{FM04})\footnote{For the comparison with functional relations in chiral Potts model, the $T^{(j)}$-fusion relations and $Q_R$-functional equation of eight-vertex model in \cite{FM04} are changed to the equivalent forms here by using the transformation, $v \mapsto v + 2 \widetilde{\eta}$.}:
\bea(cl)
T^{(j)}(v + 2 \widetilde{\eta} ) T^{(2)}(v + 2j \widetilde{\eta} ) = h^L ( v + (2j+1) \widetilde{\eta} )  T^{(j-1)}(v + 2 \widetilde{\eta})  + T^{(j+1)}( v + 2 \widetilde{\eta} )  , & 1 \leq j \leq N \ ,  \\
T^{(N+1)}(v + 2 \widetilde{\eta} )= T^{(N-1)}( v + 4 \widetilde{\eta}) + 2 h^L ( v + \widetilde{\eta} ) S^{m_1} \ &
\elea(8Fus)
which, by (\req(8TQ)), are compatible with the $T^{(j)}Q$-relations ((3.7) in \cite{FM04}):
\be
T^{(j)}(v + 2 \widetilde{\eta} )=  \sum_{m=0}^{j-1} h^L(v+(2m+1)\widetilde{\eta} )Q_R(v) Q_R^{-1}(v+2m\widetilde{\eta} )Q_R(v + 2j \widetilde{\eta} )  Q_R^{-1}(v+2(m+1) \widetilde{\eta} ) \ .
\ele(8Tjq)
Furthermore, they proposed the conjectured $Q$-functional equation ((3.2) in \cite{FM04}): 
\be
Q_L (v - {\rm i}K') = e^{\frac{\pi {\rm i} L v}{2K}} Q_L(v_0)AQ_R(v_0) \sum_{m=0}^{N-1} h(v+(2m+1) \widetilde{\eta})^L Q_R^{-1} (v + 2m \widetilde{\eta} ) Q_R (v) Q_R^{-1} (v + 2(m+1) \widetilde{\eta} ) \ ,
\ele(8QQ)
which is equivalent to the following one ((3.10) in  \cite{FM04}):
\be
 Q_R(v) \bigg( e^{\frac{\pi {\rm i} L v}{2K}} Q_L(v_0)AQ_R(v_0)\bigg)^{-1} Q_L(v - {\rm i} K') = T^{(N)}(v + 2 \widetilde{\eta} ) S^{m_1}
\ele(8TTN)
Then the ($\tau^{2}T$, fusion, $\tau^{j}T$, $(T\widehat{T})_N$, ${\sc Q}_{cp}$-functional) relations (\req(tauTSi)) (\req(SFus)) (\ref{taujTSi}) (\ref{TTNS}) and (\ref{QS}) in CPM can be matched to the corresponding ones in eight-vertex model, (\req(8TQ)), (\req(8Fus)),(\req(8Tjq)), (\req(8TTN)) and (\req(8QQ)), respectively in an identical manner by paring the  quantities in the following table: 
\bea(|c|| c  c  c |  )
\hline
&{\rm Chiral \ \ Potts \ \ model } &  & {\rm Eight \ \ vertex \ \ model}   \\
\hline
{\rm U} &U & \Longleftrightarrow  & v \mapsto v + 2 \widetilde{\eta}
\\
\hline
{\rm V} &C & \Longleftrightarrow  & v \mapsto v - {\rm i} K' 
\\
\hline
{\bf t}^{(2)}({\rm w})&\tau^{(2)} (t) & \Longleftrightarrow  & \frac{T^{(2)}( v + 2 \widetilde{\eta} )}{h^L(v+ \widetilde{\eta} ) h^L(v+ 3 \widetilde{\eta})} \\
\hline
{\bf Q}({\rm w})&{\sc Q}_{cp} (q) & \Longleftrightarrow  & Q_R (v) \\
\hline
\varphi({\rm w}) &(1- \eta^{-1} t_q) & \Longleftrightarrow  & h ( v + \widetilde{\eta})^{-1} \\
\hline
{\bf t}^{(j)}( {\rm w}) &\tau^{(j)} (t) & \Longleftrightarrow  & \frac{T^{(j)}( v + 2 \widetilde{\eta} )}{\prod_{k=1}^j h^L(v+ (2k-1) \widetilde{\eta} )} \\
\hline 
\widehat{\bf Q}({\rm w}) &\widehat{\sc Q}_{cp} (q) & \Longleftrightarrow  & Q_L (v) \\
\hline
A &\omega^L X  & \Longleftrightarrow  & I \\
\hline
B &I & \Longleftrightarrow  & S^{m_1} \\
\hline
M_j &\frac{(1- \eta^{-N} t_q^N)^L}{N^L } & \Longleftrightarrow  & e^{\frac{\pi {\rm i} L (v + 2j  \widetilde{\eta})}{2K}} Q_L(v_0)AQ_R(v_0) \\
\hline
\elea(8CP)
%$h(v - iK') =  (-1) q^{-1/2} e^{\pi {\rm i} v/K}h(v)$
Furthermore, by (\req(8CP)) one arrives the following conjectured relation in eight-vertex model corresponding to the set of $T\widehat{T}$-relations (\ref{TTS}) in CPM:
\be
 Q_R(v) \bigg( e^{\frac{\pi {\rm i} L (v + 2j  \widetilde{\eta}) }{2K}} Q_L(v_0)AQ_R(v_0)\bigg)^{-1} Q_L(v - {\rm i} K' + 2j \widetilde{\eta}) = T^{(j)}(v + 2 \widetilde{\eta} ) + T^{(N-j)}( v + 2 (j+1) \widetilde{\eta})S^{m_1} 
\ele(8TT)
for $0 \leq j \leq N$. By using (\req(8Tjq)), each relation in (\req(8TT)) is equivalent to the functional equation (\req(8QQ)) as in the CPM case.

By the comparison made above on superintegrable CPM and eight-vertex model for roots of unity, the discussion suggests that the set of functional equations might possibly occur as well for some general models in the following manner, by which when applying to CPM and eight-vertex model, the procedure produces all the functional equations in previous discussions. Consider a solvable lattice model with Boltzmann weights in a spectral curve $W$, which is ${\goth W}$ in (\ref{rapidC}) (or $W_{k'}$ in (\req(Wk'))) in CPM , and the elliptic curve with nome $q = e^{\frac{- \pi K'}{K}}$ in eight-vertex model.
The theory is built-up from families of operators of "quantum space" $\stackrel{L}{\otimes} \CZ^m$, $\{ {\bf t}^{(2)}({\rm w}), {\bf Q}  ({\rm w}), \widehat{\bf Q}({\rm w}) \}_{{\rm w} \in W}$,  such that  ${\bf Q}  ({\rm w})$ and $\widehat{\bf Q}({\rm w})$ are non-degenerated with  $ \widehat{\bf Q}({\rm w}){\bf Q}({\rm w}')= \widehat{\bf Q}({\rm w}'){\bf Q}({\rm w})$, and the following ${\bf t}{\bf Q}$-relations hold:
\bea(l)
{\bf t}^{(2)}({\rm w}) {\bf Q}  ({\rm Uw}) = \varphi ({\rm w})^L {\bf Q}  ({\rm w}) + \varphi ({\rm U w})^L {\bf Q}  ({\rm U^2 w}) A \ , \\
\widehat{\bf Q} ({\rm Uw}) {\bf t}^{(2)}({\rm w})  = \varphi ({\rm w})^L \widehat{\bf Q} ({\rm w}) + \varphi ({\rm U w})^L \widehat{\bf Q} ({\rm U^2 w}) A .
\elea(tQg)
Here $\varphi$ is a rational function on $W$, ${\rm U}$ is an automorphism of $W$ with the following "$N$-periodic" property with respect to ${\bf t}^{(2)}$ and ${\bf Q}, \widehat{\bf Q}$:
$${\bf t}^{(2)}({\rm U}^N {\rm w}) = {\bf t}^{(2)}({\rm w}) , \ \ \ {\bf Q} ({\rm U}^N {\rm w}) = {\bf Q}  ({\rm w})B , \ \ \  \widehat{\bf Q} ({\rm U}^N {\rm w}) = \widehat{\bf Q}  ({\rm w})B ,
$$ 
and $A, B$ are operators of $\stackrel{L}{\otimes} \CZ^m$, commuting with all ${\bf Q}({\rm w})$ and $\widehat{\bf Q}({\rm w})$, with $[A, B]=0$, $A^N =B^2 = 1$. Define the operators ${\bf t}^{(j)}({\rm w})$ ~ $(2 \leq j \leq N+1)$ by the ${\bf t}^{(j)}{\bf Q}$-relation:
\begin{eqnarray}
{\bf t }^{(j)}({\rm w}) 
=   \sum_{m=0}^{j-1} \bigg( \frac{\prod_{ k=0}^{j-1} \varphi({\rm U}^k{\rm w})^L}{\varphi({\rm U}^m {\rm w})^L} {\bf Q} ({\rm w})
{\bf Q}({\rm U}^m {\rm w})^{-1} {\bf Q}({\rm U}^j {\rm w}) {\bf Q}({\rm U}^{m+1}{\rm w})^{-1} A^{j-m-1} \bigg)  .
\label{tjQg}
\end{eqnarray}
Then ${\bf t}^{(j)}$'s satisfy the fusion relations by setting ${\bf t}^{(0)}=0$ and ${\bf t}^{(1)}=1$:
\bea(cll)
{\bf t}^{(j)}({\rm w}) {\bf t}^{(2)}({\rm U}^{j-1}{\rm w}) &= \varphi({\rm U}^{j-1}{\rm w})^{2L} ~  {\bf t}^{(j-1)}(t) ~ A ~ + {\bf t}^{(j+1)}({\rm w})  , & 1 \leq j \leq N \ ,  \\
{\bf t}^{(N+1)}({\rm w}) & = \varphi({\rm w})^{2L} ~ {\bf t}^{(N-1)}( {\rm U} {\rm w}) A  ~ + 2 \bigg(\prod_{k=0}^N \varphi({\rm U}^k{\rm w})\bigg)^L B \ . &
\elea(Fusg)
The fusion matrices ${\bf t}^{(j)}$'s are expected to relate to ${\bf Q}$-operator by the following ${\bf Q}\widehat{\bf Q}$-relation,
\begin{eqnarray}
{\bf Q}({\rm w}) M_j^{-1} \widehat{\bf Q}( {\rm V}{\rm U}^j {\rm w})  =&   \frac{1}{\prod_{m=0}^{j-1} \varphi({\rm U}^m {\rm w})^L} {\bf t}^{(j)}({\rm w})   + \frac{  A^j B}{\prod_{m=j}^N \varphi({\rm U}^m {\rm w})^L}{\bf t}^{(N-j)} ({\rm U}^j {\rm w}) \ , \ \ 0 \leq j \leq N , \label{TTg} 
\end{eqnarray}
where ${\rm V}$ is an order 2 automorphism of $W$, and $M_j$ are operators of $\stackrel{L}{\otimes} \CZ^m$, such that by using (\ref{tjQg}), all relations in (\ref{TTg}) reduce to a single ${\bf Q}$-functional equation:
\begin{eqnarray}
\widehat{\bf Q}( {\rm V  w} ) = M_0   \sum_{m=0}^{N-1} \frac{A^{-1-m} }{\varphi({\rm U}^m {\rm w})^L}  {\bf Q} ({\rm U}^m {\rm w})^{-1}{\bf Q} ({\rm w}) {\bf Q} ({\rm U}^{m+1}{\rm w})^{-1} .
\label{TTg}
\end{eqnarray}
The equivalent notions with those in CPM and eight vertex model are given in table (\req(8CP)). Then the functional relations among ${\bf t}^{(j)}$, ${\bf Q}$ and $\widehat{\bf Q}$ give rise to the corresponding ones in these two models. Note that among all these relations, ${\bf t}{\bf Q}$-relation (\req(tQg)) and ${\bf Q}\widehat{\bf Q}$-relation (\ref{TTg}) are the basic ones for the theory as indicated in the study of the CPM, however is still unknown for ${\bf Q}\widehat{\bf Q}$-relation in eight-vertex model case.

\section{Onsager Algebra Symmetry of $\tau^{(j)}$-matrices}
The Onsager algebra is the infinite-dimensional Lie algebra with a basis $\{ A_m, G_l \}_{ m \in \ZZ , l \in \ZZ_{>0} }$ satisfying the commutation relations:
$$
[ A_m , A_n ] = 4 G_{m-n} \ , \ \ \ [ A_m , G_l ] = 2 A_{m-l} - 2 A_{m+l} \ , \ \ \ [G_m , G_l ] = 0 \ ,
$$
where $G_{-l} := - G_l $ and $G_0 : = 0$. The elements $A_0 , A_1$ satisfy the Dolan-Grady (DG) relation \cite{DG}:
\be
[ A_1 , [ A_1 , [A_1 , A_0 ]]] = 16 [A_1 , A_0 ] \ , \ \ \ [ A_0 , [ A_0 , [A_0 , A_1 ]]] = 16 [A_0 , A_1 ] \ ,
\ele(DG)
and the Onsager algebra is characterized as the Lie algebra generated by the DG pair $\{ A_0, A_1 \}$ \cite{D91, R91}. A useful realization of Onsager algebra to identify it with the Lie-subalgebra of the $sl_2$-loop algebra, $sl_2 [z, z^{-1}]$, fixed by a standard involution through the identification \cite{R91}:
$$
A_m = 2 z^m e^+ + 2 z^{-m} e^- \ , \ \ G_m = (z^m - z^{-m}) h \ , \ \ \ \ m \in \ZZ ,
$$
where $e^\pm , h$ are $sl_2$-generators  with $[e^+, e^-]=h$, $[h, e^{\pm}] = \pm 2 e^\pm $. It is known that all finite-dimensional irreducible representations of $sl_2 [z, z^{-1}]$ are given by tensoring a finite number of irreducible $sl_2$-representations through the evaluation  of $z$ at distinct non-zero complex values $a_j$'s \cite{C}. The Hermitian irreducible representations of Onsager algebra are obtained by passing through $sl_2 [z, z^{-1}]$-representations, with the further constraints on evaluated values $a_j$'s :  
\be
| a_j | = 1, \ \ \ a_j \neq \pm 1 ,  \ \ \ a_j \neq a_k^{\pm 1} \ \ {\rm for } \ j \neq k. 
\ele(OSc)
By this,  
$A_m = 2 \sum_{j=1}^n ( a_j^m e^+_j +  a_j^{-m} e^-_j)$, $ G_m = \sum_{j=1}^n (a_j^m - a_j^{-m}) h_j$. The $a_j$'s give rise to the reciprocal polynomial of degree $2n$,
$$
\sum_{k=0}^{2n} \alpha_k z^k := \prod_{j=1}^n (z-a_j)(z-a_j^{-1})  ,
$$ 
characterized by the finite recurrence condition of $A_m$'s: $\sum_{k=0}^{2n} \alpha_k A_{k + \ell} = 0$ for $ \ell \in \ZZ$ \cite{D90, DR, R91}.

It is known that the logarithmic derivative of transfer matrices $T_p(x_q, y_q)$ at the superintegrable $p$ in (\req(Sip)) gives rise to 
the following Hamiltonian of   
$\ZZ_N$-symmetric quantum chain in \cite{GR, HKN} (see, e.g. \cite{AMP, AMPT, B88, P} and references therein),
\be
H(k') = H_0 + k' H_1
\ele(SCPM)
where $H_0$ and $H_1$ are Hermitian operators of $\stackrel{L}{\otimes}\CZ^N$ defined by
$$
H_0 = - 2\sum_{\ell=1}^L \sum_{n=1}^{N-1}
\frac{ X_\ell^n }{1-\omega^{-n}} \ , \ \ H_1 = - 2 \sum_{\ell=1}^L \sum_{n=1}^{N-1}
\frac{Z_\ell^nZ_{\ell+1}^{-n}}{1-\omega^{-n}} \ , \ \ \ \ \ ( Z_{L+1} := Z_1 ) \ . 
$$
Note that for $N=2$, this is the Ising quantum chain. The pair of operators
$$
A_0 = -2N^{-1} H_0 \ , \ \ A_1 = -2N^{-1} H_1
$$
satisfy the DG condition (\req(DG)), hence gives rise to a Hermitian representation of Onsager algebra. By this, one can study the eigenvalue problem of $H(k')$ in (\req(SCPM)) through the representation theory of Onsager algebra.    
By \cite{AMP, B88, B94} (or see the discussion in section 4 of this paper), only spin-$\frac{1}{2}$ ~ $sl_2$-representations occur in the associated Onsager algebra representations, hence $H(k')$ has the Ising-like eigenvalues:
\be
\alpha + \beta k'  + N \sum_{j=1}^n \pm \sqrt{1 + 2 k' c_j + k'^2} \ , \ \ \ \ c_j = \cos \theta_j  = \frac{1}{2}( a_j + a_j^{-1}) \ , \ \ \ \alpha , \beta \in \RZ \ .
\ele(Hk'ev)
For later use, we define
\be
\Xi_\ell = \sum_{n=1}^{N-1}
\frac{ X_\ell^n }{1-\omega^{-n}} \ , \ \ \Psi_\ell = \sum_{n=1}^{N-1}
\frac{Z_\ell^nZ_{\ell+1}^{-n}}{1-\omega^{-n}} \ , \ \ \ 1 \leq \ell \leq L \ .
\ele(avXZ)
Then $H_0 = -2 \sum_{\ell=1}^L \Xi_\ell $, $H_1 = -2 \sum_{\ell=1}^L \Psi_\ell$.

\begin{proposition}\label{prop:H0tau}
$H_0$ commutes with $\tau^2 (t)$ for all $t \in \CZ$. 
\end{proposition}
{\it Proof.} By $H_0 = -2 \sum_{\ell=1}^L \Xi_\ell $, the commutativity of $H_0$ and $\tau^2 (t_q)$ follows from $[\Xi_\ell , \tau^2 (t_q)] = 0 $ for $1 \leq \ell \leq L$. By (\req(tau2Si)), it suffices to show the commutativity of $\Xi_\ell$ and $(G_j)_\alpha^\beta (t)$ for all $\ell, j, \alpha, \beta$. By the expressions of $\Xi_\ell$ in (\req(avXZ)) and $G_\alpha^\beta (t) $ in (\req(Gsi)), one needs only to show $\Xi_\ell$ commutes with $(G_\ell)_0^1 (t) $ and $(G_\ell)_1^0 (t) $. By $Z_\ell X_\ell = \omega X_\ell Z_\ell$ and $X_\ell^N = 1$, one has
$$
\begin{array}{lll}
(1- \omega X_\ell) Z_\ell \Xi_\ell & = ( \sum_{n=0}^{N-1} \frac{\omega^n X_\ell^n}{1-\omega^{-n}} ) (1- \omega X_\ell) Z_\ell &= \bigg( \Xi_\ell  + \sum_{n=0}^{N-1} \omega^n X_\ell^n \bigg) (1- \omega X_\ell) Z_\ell  \\
& = \Xi_\ell (1- \omega X_\ell) Z_\ell ; \\
(1-  X_\ell) Z_\ell^{-1} \Xi_\ell & = ( \sum_{n=0}^{N-1} \frac{\omega^{-n} X_\ell^n}{1-\omega^{-n}} ) (1-  X_\ell) Z_\ell^{-1} &= \bigg( \Xi_\ell  - \sum_{n=0}^{N-1} X_\ell^n \bigg) (1-  X_\ell) Z_\ell^{-1} \\
& = \Xi_\ell (1- X_\ell) Z_\ell^{-1} .
\end{array}
$$
Therefore $[(G_\ell)_0^1(t), \Xi_\ell] = [(G_\ell)_1^0 (t) , \Xi_\ell] = 0$. $\Box$ \par \vspace{.2in} \noindent
\begin{proposition}\label{prop:H1tau}
$H_1$ commutes with $\tau^2 (t)$ for all $t \in \CZ$. 
\end{proposition}
{\it Proof.} By $H_1 = -2 \sum_{\ell=1}^L \Psi_\ell $, the commutativity of $H_1$ and $\tau^2 (t_q)$ follows from $[\Psi_\ell , \tau^2 (t_q)] = 0 $ for $1 \leq \ell \leq L$. For a given $\ell$, $\Psi_\ell$ commutes with $(G_j)_\alpha^\beta (\omega t)$ for $j \neq \ell, \ell+1$ in the definition (\req(tau2Si)) of $\tau^2 (t)$. Here we use the identification $L+1 = 1$. Denote by $G_{\ell, \ell+1} (t)$ the 2-by-2 matrix with the entries, $(G_{\ell, \ell+1})_\alpha^\beta$ ~ $(\alpha, \beta = 0, 1)$, defined by $(G_{\ell, \ell+1})_\alpha^\beta (t):= \sum_{j=0, 1} \bigg( (G_\ell)_\alpha^j (G_{\ell+1})_j^\beta \bigg) (t)$.  Hence it suffices to show the commutativity of $\Xi_\ell$ and $(G_{\ell, \ell+1})_\alpha^\beta (t)$ for all $ \alpha, \beta$. Using (\req(Gsi)), one has the following expressions of $(G_{\ell, \ell+1})_\alpha^\beta$:
$$
\begin{array}{cll}
(G_{\ell, \ell+1})_0^0 (t)&=& 1- \omega \eta^{-1}t X_\ell X_{\ell+1} Z_\ell Z_{\ell+1}^{-1} + \eta^{-2}t^2 X_\ell X_{\ell+1} \\
&&- \eta^{-1}t \bigg(X_\ell (1-\omega Z_\ell Z_{\ell+1}^{-1}) + X_{\ell+1}(1- Z_\ell Z_{\ell+1}^{-1}) \bigg), \\
\eta^{\frac{1}{2}}(G_{\ell, \ell+1})_0^1 (t)&= & Z_{\ell+1} - \omega^2 X_\ell X_{\ell+1} Z_\ell- \eta^{-1}t (Z_\ell - \omega X_\ell X_{\ell+1} Z_{\ell+1}) \\
&& - \bigg( \eta^{-1}t X_\ell (1-\omega Z_\ell Z_{\ell+1}^{-1}) + \omega X_{\ell+1}(1- Z_\ell Z_{\ell+1}^{-1})   \bigg) Z_{\ell+1} , \\
-\eta^{\frac{1}{2}}t^{-1}(G_{\ell, \ell+1})_1^0 (t)&= & Z_\ell^{-1} - \omega  X_\ell X_{\ell+1} Z_{\ell+1}^{-1} - \eta^{-1}t (Z_{\ell+1}^{-1} - X_\ell X_{\ell+1} Z_\ell^{-1}) \\
&& - \bigg(X_\ell (1-\omega Z_\ell Z_{\ell+1}^{-1}) - X_{\ell+1}(1- Z_\ell Z_{\ell+1}^{-1}) \bigg) Z_\ell^{-1} , \\
(G_{\ell, \ell+1})_1^1 (t)&=& \omega^2 X_\ell X_{\ell+1} - \eta^{-1}t (1+ \omega X_\ell X_{\ell+1}) Z_\ell^{-1} Z_{\ell+1} + \eta^{-2}t^2 \\
&& - \bigg(X_\ell (1-\omega Z_\ell Z_{\ell+1}^{-1}) + \omega X_{\ell+1}(1- Z_\ell Z_{\ell+1}^{-1}) \bigg) Z_\ell^{-1} Z_{\ell+1} .
\end{array}
$$
By the expression of $\Psi_\ell$ in (\req(avXZ)), $\Psi_\ell$ commutes with $Z_\ell, Z_{\ell+1}$ and $X_\ell X_{\ell+1}$. Furthermore, by 
$$
\begin{array}{ll}
\Psi_\ell X_\ell (1 - \omega Z_\ell Z_{\ell+1}^{-1}) & =
X_\ell (\Psi_\ell + \sum_{n=0}^{N-1} \omega^n Z_\ell^nZ_{\ell+1}^{-n})(1 - \omega Z_\ell Z_{\ell+1}^{-1})= X_\ell (1 - \omega Z_\ell Z_{\ell+1}^{-1}) \Psi_\ell ,  \\
\Psi_\ell X_{\ell+1} (1 - Z_\ell Z_{\ell+1}^{-1} ) &= 
X_{\ell+1} (\Psi_\ell - \sum_{n=0}^{N-1} Z_\ell^nZ_{\ell+1}^{-n})(1 - Z_\ell Z_{\ell+1}^{-1} ) = X_{\ell+1} (1 - Z_\ell Z_{\ell+1}^{-1} ) \Psi_\ell ,
\end{array}
$$
all the terms appeared in the expressions of $(G_{\ell, \ell+1})_\alpha^\beta$ commute with $\Psi_\ell$. Then follows $[\Psi_\ell , \tau^2 (t_q)] = 0$.
$\Box$ \par \vspace{.2in} 
Since $H_0, H_1$ form the DG pair which generates the Onsager algebra representations, Proposition \ref{prop:H0tau} and \ref{prop:H1tau} imply the Onsager algebra symmetry of $\tau^{(2)}(t)$. Note that the spin-shift operator $X ( = \prod_{\ell = 1}^L X_\ell)$ commutes with $H_0$ and $H_1$.
By the expression (\ref{SitauF}) of $\tau^{(j)}$, we obtain the following result:
\begin{theorem}\label{thm:H0tau}
 The $\tau^{(j)}$-operators in the superintegrable CPM possess the Onsager algebra symmetry, i.e.,  $\tau^{(j)}(t)$'s commute with $H_0$ and $H_1$, hence the corresponding Onsager irreducible representations give rise to the degenerated eigenvectors of $\tau^{(j)}(t)$ for $2 \leq j \leq N$. Indeed, the eigenvectors of $H(k')$ associated to the eigenvalues $(\req(Hk'ev))$ form a degenerated eigenspace of $\tau^{(j)}(t)$'s. 
\end{theorem} $\Box$ \par \vspace{.1in} \noindent
{\bf Remark.} By the commutative property (\req(tau2T)) of $\tau_p^{(2)}$ and $T_p$ for the superintegrable element $p$, $\tau^{(2)}$ commutes with the Hamiltonian $H (k')$ in (\req(SCPM)). Since $H(k')$ is the sum of $H_0$ and $k'H_1$ for a non-zero $k'$ which is related to $\eta$ in $\tau^{(2)}$ by $\eta= ( \frac{1-k'}{1+k'})^{\frac{1}{N}}$, Proposition \ref{prop:H0tau} is equivalent to Proposition \ref{prop:H1tau} when employing the theory of CPM. Here we give the algebraic verification of Onsager algebra symmetry of $\tau^{(2)}(t)$, hence $\tau^{(j)}$-matrices, directly from the explicit form of $\tau^{(2)}$-matrices and Onsager algebra generators, not making use of the commutative relation (\req(tau2T)) of $\tau^{(2)}$ and $T$. However, to questions concerning the nature of the representations of Onsager algebra, one still needs to use the complete theory of superintegrable CPM, which we are going to discuss in the next section.
$\Box$ \par \vspace{.2in} \noindent

\section{Bethe Equation of $\tau^{(2)}$-matrix, Degeneracy and Eigenvalues of $\tau^{(j)}$-matrix}
In this section, we discuss in details the nature of Onsager algebra representations and $\tau^{(j)}$-eigenvalues  in Theorem \ref{thm:H0tau} using results essentially taken from works in \cite{AMPT9, AMP, B93, B94}. 

In the discussion of superintegrable CPM in section 1, it is natural to introduce the variable
$$
\widetilde{t}= \eta^{-1} t 
$$
where as before, $\eta= (\frac{1-k'}{1+k'})^{\frac{1}{N}}$. 
The hyperelliptic curve (\req(Wk')) becomes 
\be
\widetilde{t}^N = \frac{(1- k' \lambda)(1 - k' \lambda^{-1})}{(1- k')^2 } \ , \ \ \ {\rm or \ equivalently}, \ \ \ \ \ \ (\frac{\lambda  + 1}{\lambda   - 1})^2 = \frac{ \widetilde{t}^N - \eta^{-2N}  }{\widetilde{t}^N - 1} \ , 
\ele(Wk'-)
i.e., $ 
 \bigg( \frac{ \widetilde{t}^N - \eta^{-2N}  }{\widetilde{t}^N - 1}\bigg)^{\frac{1}{2}}  = \frac{\lambda^{\pm 1}  + 1}{\lambda^{\pm 1} - 1}$. Apply the gauge transform $M^{-1}G(t) M$ to $G(t)$ in (\req(Gsi)) with $ M = {\rm dia}. [1,\eta^{\frac{1}{2}}]$, then $M^{-1}G(t) M$ has the expression
$$
\widetilde{G} (\widetilde{t})  = \left( \begin{array}{cc}
       1 - \widetilde{t} X & ( 1 - \omega  X )  Z   \\
       - \widetilde{t} (1 - X ) Z^{-1} & - \widetilde{t}   + \omega   X  
\end{array} \right)   ,
$$
which is again a Yang-Baxter solution (\req(G)) with $a=b=c=d=1$. Then $\tau^{(2)}(t)$ in (\req(tau2Si)) can be written as  $\tau^{(2)}(t)= \widetilde{\tau}^{(2)}(\widetilde{t})$, where 
\be
\widetilde{\tau}^{(2)}(\widetilde{t}) = {\rm tr}_{aux} ( \bigotimes_{\ell=1}^L \widetilde{G}_\ell) ( \omega \widetilde{t}) \ , \ \ \ \ \ G_\ell (\widetilde{t}) = G(\widetilde{t}) \ {\rm at \ \ell th \ site}. 
\ele(tau2-)
Write $\tau^{(j)}(t)= \widetilde{\tau}^{(j)}(\widetilde{t})$, then the fusion relation (\req(SFus)) becomes  
\bea(cll)
\widetilde{\tau}^{(j)}(\widetilde{t}) \widetilde{\tau}^{(2)}(\omega^{j-1} \widetilde{t}) &= (1 - \omega^{j-1} \widetilde{t} )^{2L} ~  \widetilde{\tau}^{(j-1)}(\widetilde{t}) ~ \omega^L X ~ + \widetilde{\tau}^{(j+1)}(\widetilde{t})  , & 1 \leq j \leq N \ ,  \\
\widetilde{\tau}^{(N+1)}(\widetilde{t}) & = (1 - \widetilde{t} )^{2L} ~ \widetilde{\tau}^{(N-1)}( \omega \widetilde{t}) \omega^L X  ~ + 2 (1 - \widetilde{t}^N  )^L  \ . &
\elea(Fus-) 
Note that the $\widetilde{\tau}^{(j)}$-matrices are temperature ( i.e. $k'$) independent. 
With operators ${\sc Q}_{cp}$, $ \widehat{\sc Q}_{cp}$ in (\req(Qcp)), the relations (\req(tauTSi))-(\ref{TTNS}) become the following $\widetilde{\tau}^{(2)}{\sc Q}_{cp}$, $\widetilde{\tau}^{(j)}{\sc Q}_{cp}$-relations,
\be
\widetilde{\tau}^{(2)}(\omega^{-1}\widetilde{t}_q) {\sc Q}_{cp} (q) = (1 - \omega^{-1}\widetilde{t}_q)^L {\sc Q}_{cp} (U^{-1}q) + (1 - \widetilde{t}_q)^L  {\sc Q}_{cp} (Uq) \omega^L X \ ,  
\ele(tauT-)
\be
\widetilde{\tau}^{(j)}(\widetilde{t}_q) 
=  \sum_{m=0}^{j-1} \bigg( \frac{\prod_{ k=0}^{j-1} (1- \omega^k \widetilde{t}_q)^L}{(1- \omega^m \widetilde{t}_q)^L} {\sc Q}_{cp}(q)
{\sc Q}_{cp}(U^m q)^{-1} 
{\sc Q}_{cp}(U^j q) {\sc Q}_{cp}(U^{m+1}q)^{-1} (\omega^L X)^{j-m-1} \bigg)  ,
\ele(taujT-)
and the ${\sc Q}_{cp}\widehat{\sc Q}_{cp}$-relation
\begin{eqnarray}
&{\sc Q}_{cp}(q) \widehat{\sc Q}_{cp}( C U^j q )  = \frac{(1-  \widetilde{t}_q^N)^L}{N^L }  \bigg( \frac{1}{\prod_{m=0}^{j-1} (1 - \omega^m \widetilde{t}_q)^L } \widetilde{\tau}^{(j)}(\widetilde{t}_q)   + \frac{ (\omega^L X)^j }{\prod_{m=j}^{N-1} (1 - \omega^m \widetilde{t}_q)^L} \widetilde{\tau}^{(N-j)} (\omega^j \widetilde{t}_q)  \bigg), & 0 \leq j \leq N ; \nonumber \\
&{\sc Q}_{cp}(q) \widehat{\sc Q}_{cp}( Cq )  = \frac{1}{N^L} \widetilde{\tau}^{(N)}(\widetilde{t}_q) \ . 
\label{TT-}
\end{eqnarray}

By \cite{AMP, B93, B94}, eigenvalues of the transfer matrix of CPM in the
superintegrable case are solved by Bethe-ansatz method as follows. Let $F( \widetilde{t})$ be a $\widetilde{t}$-polynomial (of degree
$m_p$): 
\be
F(\widetilde{t}) ~ ( =  F (\widetilde{t} ;
v_1, \ldots, v_{m_p} ) ) = \prod_{i =1}^{m_p} ( 1 + \omega v_i \widetilde{t} )
\ele(F)
where $v_1, \ldots, v_{m_p}$ are the parameters with $(-v_i )^N \neq 0, 1$ and $v_i v_j^{-1} \neq 1, \omega$ for $i \neq j$. Associated to $F(\widetilde{t})$, we consider the following ${\cal P}$-function  
\be
{\cal P} ( \widetilde{t} ) ~ ( = {\cal P} ( \widetilde{t} ; v_1, \ldots, v_{m_p}) )  =  \omega^{-P_b}
\sum_{j=0}^{N-1}
\frac{(1-\widetilde{t}^N)^L(\omega^j\widetilde{t})^{-P_a-P_b} }{(1-
\omega^j \widetilde{t})^L F(\omega^j \widetilde{t}) F(
\omega^{j+1} \widetilde{t})} \ ,
\ele(ptv)
where $P_a, P_b$ are integers 
satisfying\footnote{Here we consider only the case with periodic boundary condition  by assuming $r$ in \cite{B93, B94} equal to $0$.}  
\be
0 \leq P_a+P_b \leq N-1, ~ \ \ \ \ P_b-P_a \equiv Q+L \pmod{N}, 
\ele(Pab)
with $Q$ the $\ZZ_N$-charge as before. For 
arbitrary $v_1, \ldots, v_{m_p}$,
${\cal P}( \widetilde{t} )$ is a rational function invariant under
the transformation: $\widetilde{t}
\mapsto \omega \widetilde{t}$. The necessary and sufficient condition
for ${\cal P} ( \widetilde{t} )$ to be a $\widetilde{t}$-polynomial is the Bethe-ansatz-type constraint for parameters $v_1, \ldots, v_{m_p}$, ((4.4) in \cite{AMP}, (6.22) in \cite{B93}):
\be
( \frac{v_i + \omega^{-1}}{v_i+ \omega^{-2}} )^L
= - \omega^{-P_a-P_b} \prod_{l=1}^{m_p}
\frac{v_i - \omega^{-1}v_l}{v_i - \omega v_l} \ ,
\ \ \ i = 1, \ldots, m_p \ .
\ele(CPMBe)
The above relation, up to the phase factor,  is the Bethe equation of spin-$\frac{N-1}{2}$ XXZ chain for size $L$ with the anisotropy $\gamma = \frac{\pi}{N}$ and periodic boundary condition \cite{B93, DKM}.  An equivalent form to the above Bethe equation is
\be
( \frac{v_i + \omega^{-1}}{v_i+ \omega^{-2}} )^L
= - \omega^{L+Q-2P_b} \prod_{l=1}^{m_p}
\frac{v_i - \omega^{-1}v_l}{v_i - \omega v_l} , \ \ \ i = 1, \ldots, m_p \ ,
\ele(Be')
and $P_a$ is the integer determined by (\req(Pab)). 
Then ${\cal P}( \widetilde{t} )$ is a $\widetilde{t}^N$-polynomial of degree $m_E$ with zeros $\widetilde{t}^N = s_1, \ldots, s_{m_E}$. Note that by the theory of the CPM ${\cal P} (1) \neq 0$, indeed its value determines the total momentum $P$ through the relation
\be
e^{{\rm i}P}  =  \omega^{-P_b}\prod_{i =1}^{m_p} \frac{1 + \omega v_i  }{ 1 + \omega^2 v_i} = \frac{{\cal P} (1)}{N^L} \prod_{i =1}^{m_p} ( 1 + \omega v_i )^2 
\ele(P)
(see (2.24) of \cite{AMP}). By $e^{{\rm i}P L} = 1$,  the relation (\req(P)) yields the restriction of the non-negative integer $m_p$ in  (\req(Be')):
 \be
LP_b \equiv m_p (Q-2P_b-m_p) \pmod{N}, 
\ele(mp)
(for $N=3$ case, see (C.3) (C.4) in \cite{AMP}). Conversely, if $m_p$ satisfies the above relation, the right hand side of (\req(P)) is a $L$th root of unity, hence determines the total momentum $P$. Furthermore the integer $P_b$ in the theory is chosen to guarantee ${\cal P }(0) \neq 0$, (which will become clear later on). Therefore the $\widetilde{t}^N$-zeros of the polynomial ${\cal P}( \widetilde{t} )$ have all   
$s_j \neq 0 , 1 $. 
For each $j$, let $\lambda_j$ be the complex number defined by the relation (\req(Wk'-)) with $(\widetilde{t}^N, \lambda) = (s_j , \lambda_j^{\pm 1})$, and denote $\pm w_j = \frac{\lambda_j^{\pm 1} +1}{\lambda_j^{\pm 1} -1}  = (\frac{ s_j - \eta^{-2N}}{  s_j -1 })^{\frac{1}{2}} $.  Hence $\lambda_j \neq 1, k'^{\pm 1}, 0, \infty$, equivalently, 
$w_j \neq \infty, \pm \frac{k'+1}{k'-1},  \pm 1$. Define 
\be
{\cal G} (\lambda) = \prod_{j=1}^{m_E} \frac{\lambda + 1 \pm (\lambda - 1) w_j}{2\lambda} \ .
\ele(Gpo)
Then ${\cal G}(k'^{\pm 1}), {\cal G}(0)$ are non-zero. By $\frac{1}{4}(\lambda + 1 + (\lambda - 1) w_j)( \lambda^{-1} + 1 + (\lambda^{-1} - 1) w_j)= \frac{\widetilde{t}^N - s_j}{ 1-s_j  }$, one has the following relation between the ${\cal P}$-polynomial (\req(ptv)) and ${\cal G}(\lambda)$: 
\be
\frac{{\cal P}(\widetilde{t})}{{\cal P}(1)} = {\cal G} (\lambda){\cal G} (\lambda^{-1}) .
\ele(GP)
By (1.11) in \cite{AMP} and (21) in \cite{B94}, one has the following $T(q)$-eigenvalue\footnote{Here we modify the expression (21) of \cite{B94} by a multiple factor $e^{-{\rm i}P/2}$ so that $T(q)$ takes the value 1 for $q= p$, which would be more appropriate for the consistency with the analysis made in formula (1.11) in \cite{AMP}, elucidated later on in this section.  Note that the expression here agrees with the conjectured formula (2.22) of \cite{AMP}, in which the notations ${\cal N}$, $\lambda$, $\eta$, $NP_c$, $w_\ell$ correspond respectively to $L$, $k'$, $\eta^{\frac{-1}{2}}$, $P_\mu$, $\frac{1-k'}{2} w_\ell$ used in this article. Indeed the $T_q$ in formula  (21) of \cite{B94} is equal to $N^L \frac{(\eta^{\frac{-1}{2}} x_q-1)^L}{(\eta^{-N/2} x_q^N-1)^L} (\eta^{\frac{-1}{2}}x_q)^{P_a}(\eta^{\frac{-1}{2}}y_q)^{P_b}\mu_q^{-P_\mu} \frac{F(\widetilde{t}_q) }{ (\omega^{P_b} F(1) F(\omega))^{1/2} } {\cal G} (\lambda_q)$.}:  
\be
T(q) = N^L \frac{(\eta^{\frac{-1}{2}} x_q-1)^L}{(\eta^{-N/2} x_q^N-1)^L} (\eta^{\frac{-1}{2}}x_q)^{P_a}(\eta^{\frac{-1}{2}}y_q)^{P_b}\mu_q^{-P_\mu} \frac{F(\widetilde{t}_q) }{F(1)} {\cal G} (\lambda_q) ,
\ele(Tqval)
and $\widehat{T}(q)$ is equal to $ e^{{\rm i}P}T(q)$ with the expression
\be
\widehat{T}(q) = N^L \frac{(\eta^{\frac{-1}{2}} x_q-1)^L}{(\eta^{-N/2} x_q^N-1)^L} (\eta^{\frac{-1}{2}}x_q)^{P_a}(\eta^{\frac{-1}{2}}y_q)^{P_b}\mu_q^{-P_\mu} \frac{F(\widetilde{t}_q) }{ \omega^{P_b} F(\omega) } {\cal G} (\lambda_q).
\ele(hTqval)
Hence we have the following eigenvalues for operators ${\sc Q}_{cp}(q)$ and $\widehat{\sc Q}_{cp} (q)$:
\bea(lll)
{\sc Q}_{cp} (q) &= (\eta^{\frac{-1}{2}}x_q)^{P_a}(\eta^{\frac{-1}{2}}y_q)^{P_b}\mu_q^{-P_\mu} \frac{F(\widetilde{t}_q) }{F(1)} {\cal G} (\lambda_q) ,& \\
\widehat{\sc Q}_{cp} (q) &= (\eta^{\frac{-1}{2}}x_q)^{P_a}(\eta^{\frac{-1}{2}}y_q)^{P_b}\mu_q^{-P_\mu} \frac{F(\widetilde{t}_q) }{ \omega^{P_b} F(\omega) } {\cal G} (\lambda_q) &(= e^{{\rm i}P} {\sc Q}_{cp} (q)) ,
\elea(Qval)
which imply ${\sc Q}_{cp}(p) = {\bf 1}$, $\widehat{\sc Q}_{cp} (p) = e^{{\rm i}P}{\bf 1}$.
As $q$ tends to $p$, by setting $x_q = \eta^{\frac{1}{2}} ( 1 - 2 k' \varepsilon + O ( \varepsilon ^2))$ with small $\varepsilon$, to the first order one has $y_q = \eta^{\frac{1}{2}} ( 1 + 2 k' \varepsilon)$ , $\mu_q = 1 + 2 (k'-1) \varepsilon$, and formula (1.11) in \cite{AMP} for the CPM-transformation matrix
$$
T(x_q, y_q) = {\bf 1} [ 1 + \varepsilon (N-1)L ] + \varepsilon H (k') + O ( \varepsilon ^2) 
$$
where $H(k')$ is the Hamiltonian (\req(SCPM)). 
Then the $T(q)$-eigenvalue (\req(Tqval)) gives rise to the following energy value of $H(k')$ ((2.23) of \cite{AMP}):
\be
E = 2 P_\mu + N m_E - (N-1)L + k' \bigg( (N-1)L -2P_\mu-Nm_E + 2(P_b-P_a)\bigg) + N(1-k') \sum_{j=1}^{m_E} \pm w_j .
\ele(E)
Write $
H(k') = \frac{-N}{2} (A_0 + k' A_1)$ as before with
 Hermitian $(\stackrel{L}{\otimes}\CZ^N)$-operators $A_0, A_1$ satisfying the DG relation (\req(DG)), hence generating a Hermitian representation of Onsager algebra. Each Onsager algebra irreducible subrepresentation of $\stackrel{L}{\otimes} \CZ^N$ 
is obtained by an evaluation representation of $sl_2 [z, z^{-1}]$ by evaluating $z$ at finite complex numbers $a_j$'s satisfying (\req(OSc)), then composed with irreducible representations of the $sl_2$-factors, characterized by the spin $\sigma_j \in \frac{1}{2} \ZZ_{\geq 0}$. Through this, one can write $A_0 =  \frac{-N}{2} \alpha + 2 \sum_{j} (e_j^+ + e_j^-)$, $A_1 = \frac{-N}{2} \beta + 2 \sum_{j} (a_j e_j^+ + a_j^{-1} e_j^-)$, which imply the eigenvalues of $H(k')$ with the form
$$
\alpha + \beta k' + 2N \sum_{j} m_j \sqrt{1+ 2k' \cos \theta_j + k'^2}
$$
where $\alpha, \beta \in \RZ$, $\cos \theta_j = a_j + a_j^{-1}$, and $m_j$ runs through $-\sigma_j, -\sigma_j+1, \ldots, \sigma_j$ \cite{D90, D91, DR, R91}.
Comparing the above formula with the energy expression (\req(E)), we find  
\be
\sigma_j = \frac{1}{2}, \ \ \ \cos \theta_j = \frac{1+ s_j}{1- s_j } \ \ \ \ {\rm for \ all } \ j  \ ,
\ele(sppa)
then follows the expression (\req(Hk'ev)). 
Since $\cos \theta_j$'s are distinct real numbers with $| \cos \theta_j | < 1$, the $\widetilde{t}^N$-polynomial associated to ${\cal P}(\widetilde{t})$ in (\req(ptv)) is simple with negative real roots $s_j \ ( 1 \leq j \leq m_E )$. In particular, ${\cal P}(0) \neq 0$.
Hence we have shown $(i)$ of the following theorem except the formula of $m_E$.
\begin{theorem}\label{thm:Ply} (i) Let $\{ v_i \}_{i=1}^{m_p}$ be a solution of the equation $(\req(Be'))$ for $L, Q, P_b$ with the relation $(\req(mp))$, and ${\cal P}(\widetilde{t})$  the polynomial $(\req(ptv))$ associated to $\{ v_i \}_{i=1}^{m_p}$. Then ${\cal P}(0) \neq 0$, and ${\cal P}(\widetilde{t})$ is a simple $\widetilde{t}^N$-polynomial of degree $m_E$ with negative real roots. Furthermore, $m_E = [\frac{(N-1)L -P_a-P_b-2m_p}{N}]$ where $[r]$ denotes the integral part of a real number $r$. 

(ii) For $\{ v_i \}_{i=1}^{m_p}$ in (i) with the total momentum $P$, denote $v_i'= \omega^{-3}v_i^{-1}$, and $Q' \equiv -L-Q$, $P_b' \equiv -P_b-m_p  \pmod{N}$. Then $\{ v'_i \}_{i=1}^{m_p}$ is a Bethe solution of $(\req(Be'))$ for $L, Q', P_b'$ with the total momentum $P'= -P$. Furthermore, the polynomial ${\cal P}(\widetilde{t}; v_1', \ldots, v'_{m_p})$ in $(\req(ptv))$ associated to $\{ v_i' \}_{i=1}^{m_p}$ is related to ${\cal P}(\widetilde{t}; v_1, \ldots, v_{m_p})$ by the following reciprocal relation,
$$
{\cal P}(\widetilde{t}; v_1', \ldots, v'_{m_p}) \omega^{P_b'}\prod_{i =1}^{m_p}  v'_i= \widetilde{t}^{Nm_E} ~ {\cal P}(\widetilde{t}^{-1}; v_1, \ldots, v_{m_p}) \omega^{P_b} \prod_{i =1}^{m_p}  v_i  \ .
$$
\end{theorem}
{\it Proof.} It is easy to see that $m_E \leq [\frac{(N-1)L -P_a-P_b-2m_p}{N}]$. Let $v'_i, Q', P'_b, P'$ be those defined in $(ii)$ in connection with $v_i, Q, P_b, P$ in $(i)$. One can easily see that $\{v_i \}_{i=1}^{m_p}$ as a Bethe solution of $(\req(Be'))_{L, Q, P_b}$ is equivalent to that for $\{v'_i \}_{i=1}^{m_p}$ in $(\req(Be'))_{L, Q', P'_b}$; the same for $(P, \{v_i \}_{i=1}^{m_p})$ and $(P', \{v'_i \}_{i=1}^{m_p})$ with the relation (\req(P)). By (\req(Pab)),  $P_b-P_a \equiv Q+L$, $P'_b-P'_a \equiv Q'+L \pmod{N}$, and $0 \leq P_a+P_b, P'_a+P'_b \leq N-1$. Hence $P'_a +P'_b \equiv -L -P_a -P_b-2m_p \pmod{N}$, which implies 
$$
\begin{array}{l}
[\frac{-L -P_a -P_b-2m_p}{N}]= [\frac{-L -P'_a -P'_b-2m_p}{N}] =\frac{-L -P_a -P_b-2m_p -P'_a-P'_b}{N} \ ,
\end{array}
$$
and $[\frac{(N-1)L -P_a-P_b-2m_p}{N}]= \frac{(N-1)L -P_a -P_b-2m_p -P'_a-P'_b}{N}$. Using the relation
$$
\frac{(1-\widetilde{t}^{N})^L \widetilde{t}^{-P_a'-P_b'} \prod_{i =1}^{m_p}  v'_i }{(1-\widetilde{t})^L \prod_{i =1}^{m_p} \bigg( ( 1+ \omega v'_i
 \widetilde{t} ) (1+ \omega^2 v'_i   ) \bigg) }= \frac{\widetilde{t}^{(N-1)L- 2m_p- P_a-P_b-P_a'-P_b'} (1-\widetilde{t}^{-N})^L \widetilde{t}^{P_a+P_b} \prod_{i =1}^{m_p}  v_i }{(1-
\widetilde{t}^{-1})^L \prod_{i =1}^{m_p} \bigg( ( 1 + \omega v_i \widetilde{t}^{-1} )( 1 + \omega^2 v_i \widetilde{t}^{-1} )\bigg) } , 
$$
one has 
$$
{\cal P}(\widetilde{t}; v_1', \ldots, v'_{m_p}) \omega^{P_b'}\prod_{i =1}^{m_p}  v'_i= \widetilde{t}^{N[\frac{(N-1)L -P_a-P_b-2m_p}{N}]} ~ {\cal P}(\widetilde{t}^{-1}; v_1, \ldots, v_{m_p}) \omega^{P_b} \prod_{i =1}^{m_p}  v_i  \ .
$$
Set $\widetilde{t}=0$ in the above relation, then ${\cal P}(0; v_1', \ldots, v'_{m_p}) \omega^{P_b'}\prod_{i =1}^{m_p}  v'_i= C \omega^{P_b} \prod_{i =1}^{m_p}  v_i$, where $C$ is the coefficient of $\widetilde{t}^{N[\frac{(N-1)L -P_a-P_b-2m_p}{N}]}$ in the polynomial ${\cal P}(\widetilde{t}; v_1, \ldots, v_{m_p})$. By $(i)$ for ${\cal P}(0)$ (associated to $v'_i$s) , we have ${\cal P}(0; v_1', \ldots, v'_{m_p}) \neq 0$, hence $C \neq 0$. Then follow the equality $m_E = [\frac{(N-1)L -P_a-P_b-2m_p}{N}]$ in $(i)$, and the reciprocal relation in $(ii)$. 
$\Box$ \par \vspace{.1in} \noindent
{\bf Remark}. The dominant eigenvalues of $T(q)$ are in the case $P_b = m_p =0$ \cite{B94}, where $m_E ~ (= [\frac{(N-1)L-P_a}{N}])= [\frac{(N-1)L-Q}{N}]$. The mathematical structures of polynomials ${\cal P}(\widetilde{t})$ and the corresponding polynomials in Onsager algebra representations were discussed in  \cite{G02, GRo, R05i}
$\Box$ \par \vspace{.2in}

Let $\{ v_i \}_{i=1}^{m_p}$ be a Bethe solution of (\req(Be')), and $F(\widetilde{t})$ the polynomial (\req(F)) defined by  $v_i$'s.  Associated to $F(\widetilde{t})$, one has the ${\sc Q}_{cp}$and $\widehat{\sc Q}_{cp}$-eigenvalues (\req(Qval)), with the corresponding $\tau^{(j)}$-eigenvalue for all $j$. Then relations (\req(tauT-)) and (\req(Pab)) yield ((6.18) in \cite{B93}): 
$$
\begin{array}{l}
\widetilde{\tau}^{(2)}(\omega^{-1}\widetilde{t}_q)  F(\widetilde{t}_q)  = \omega^{-P_a} (1 - \omega^{-1}\widetilde{t}_q)^L   F(\omega^{-1} \widetilde{t}_q)   + \omega^{P_b} (1 - \widetilde{t}_q)^L     F(\omega \widetilde{t}_q)   \ .
\end{array}  
$$
Set $\widetilde{t}_q = - (\omega v_i)^{-1}$ in the above relation. One obtains the Bethe equation (\req(CPMBe)), equivalent to (\req(Be')), of $v_i$'s with the following polynomial expression for $\widetilde{\tau}^{(2)}$-eigenvalue:
\be
\widetilde{\tau}^{(2)}(\widetilde{t}) ~ ( = \widetilde{\tau}^{(2)}(\widetilde{t} ; v_1, \ldots, v_{m_p} ) )  =  \frac{\omega^{-P_a-P_b} (1 - \widetilde{t})^L  \prod_{i =1}^{m_p} ( 1 + \omega v_i \widetilde{t} )    +  (1 - \omega \widetilde{t})^L \prod_{i =1}^{m_p} ( 1 + \omega^3 v_i \widetilde{t} )  }{\omega^{-P_b}\prod_{i =1}^{m_p} ( 1 + \omega^2 v_i \widetilde{t} ) }.
\ele(t2-v)
By (\ref{taujT-}), one has the $\widetilde{\tau}^{(j)}$-polynomial expression:
\be
\widetilde{\tau}^{(j)}(\widetilde{t}) =  \omega^{(j-1)P_b} \prod_{ k=0}^{j-1} (1- \omega^k \widetilde{t})^L  \sum_{m=0}^{j-1}  \frac{ F(\widetilde{t}) F(\omega^j \widetilde{t}) \omega^{-m ( P_a +P_b)}}{(1- \omega^m \widetilde{t})^L F(\omega^m \widetilde{t})F(\omega^{m+1} \widetilde{t})}, \ \ \   2 \leq j \leq N .
\ele(tjval-)
Then the following relations ((6.17) and (6.24) in \cite{B93})\footnote{The formulae here differ with those in \cite{B93} by some scalar factors as we use conventions in \cite{AMP, B94} in this article.} hold:
\bea(cl)
 \widetilde{t}^{P_a+P_b} F(\widetilde{t}) F(\omega^j \widetilde{t}) {\cal P} (\widetilde{t})  = \frac{\omega^{-jP_b}(1-  \widetilde{t}^N)^L}{\prod_{m=0}^{j-1} (1 - \omega^m \widetilde{t})^L } \widetilde{\tau}^{(j)}(\widetilde{t})   +  \frac{\omega^{-jP_a} (1-  \widetilde{t}^N)^L  }{\prod_{m=j}^{N-1} (1 - \omega^m \widetilde{t})^L} \widetilde{\tau}^{(N-j)} (\omega^j \widetilde{t})  , & 0 \leq j \leq N , \\
\widetilde{\tau}^{(N)}(\widetilde{t}) = \widetilde{t}^{P_a+P_b} F(\widetilde{t})^2 {\cal P} (\widetilde{t})  & ({\rm for} \ j=N )  
\elea(tjpols) 
where ${\cal P} (\widetilde{t})$ is in (\req(P)). Note that by (\req(GP)), the above  relation is equivalent to the ${\sc Q}_{cp}\widehat{\sc Q}_{cp}$-relation (\ref{TT-}). Hence we have shown the following results:
\begin{theorem}\label{thm:Bethe}
$(i)$ The Bethe equation of $\widetilde{\tau}^{(2)}(\omega^{-1}\widetilde{t})$ for the $L$-size $\widetilde{\tau}^{(2)}$-matrix $(\req(tau2-))$ in the six-vertex model (with external fields), is the relation $(\req(Be'))$ (or equivalent to $(\req(CPMBe))$), where the equation depends on two quantum numbers $Q, P_b \in \ZZ_N $ with the constraint $(\req(mp))$ on  non-negative integer $m_p$. The total momentum $P$ is given by $(\req(P))$.

$(ii)$ Let $\{ v_i \}_{i=1}^{m_p}$ be a Bethe solution of $(\req(Be'))$, and $F(t) (= F(t; v_1, \ldots, v_{m_p}))$ the polynomial defined in $(\req(F))$. Then  there associates an eigen-polynomial $\widetilde{\tau}^{(j)}(\widetilde{t}) ~ (= \widetilde{\tau}^{(j)}(\widetilde{t}; v_1, \ldots, v_{m_p}))$ of the $\widetilde{\tau}^{(j)}$-operator for $2 \leq j \leq N$, given by $(\req(tjval-))$ such that the relation $(\req(tjpols))$ holds.  
\end{theorem} $\Box$ \par \vspace{.1in} \noindent
{\bf Remark}. One may as well obtain polynomials $\widetilde{\tau}^{(j)}(\widetilde{t})$'s  by using the $\widetilde{\tau}^{(2)}(\widetilde{t})$-expression $(\req(t2-v))$ and the fusion relation $(\req(Fus-))$. 
$\Box$ \par \vspace{.2in} \noindent

Note that Theorem \ref{thm:Bethe} was derived from ${\sc Q}_{cp}$-eigenvalues in (\req(Qval)), where all possible ${\cal G}(\lambda_q)$'s give the same $\widetilde{\tau}^{(j)}$-polynomial for $2 \leq j \leq N$. By (\req(Gpo)) and (\req(E)), the eigen-vectors corresponding to those available ${\sc Q}_{cp}$-eigenvalues form a $2^{m_E}$-dimensional vector space which is the representation space for an irreducible Onsager algebra representation with the evaluation parameters determined by the ${\cal P}$-polynomial (\req(ptv)) via the connection (\req(sppa)). Hence the Onsager algebra describes the full symmetry of the degeneracy space for an given  $\widetilde{\tau}^{(j)}$-polynomial.
This can be regarded as an 
explicit formulation of the Onsager algebra symmetry of $\tau^{(j)}$-matrix in Theorem \ref{thm:H0tau}. As a corollary of Theorem \ref{thm:Ply} $(i)$, we have the following result.
\begin{proposition}\label{prop:OAsym}
Let $\{ v_i \}_{i=1}^{m_p}$ be a Bethe solution of $(\req(Be'))$, $\widetilde{\tau}^{(j)}(\widetilde{t}) ~ (= \widetilde{\tau}^{(j)}(\widetilde{t}); v_1, \ldots, v_{m_p})$ the eigen-polynomial in $(\req(tjval-))$ for $2 \leq j \leq N$. There is an irreducible Onsager algebra subrepresentation space (of $\stackrel{L}{\otimes} \CZ^N$) of dimension $2^{[\frac{(N-1)L -P_a-P_b-2m_p}{N}]}$ with $\widetilde{\tau}^{(j)}(\widetilde{t})$ as the $\widetilde{\tau}^{(j)}$-eigenvalue.  
\end{proposition} $\Box$ \par \vspace{.2in}

\section{Concluding Remarks}
In this paper we have shown the Onsager algebra symmetry of $\tau^{(j)}$-matrices in superintegrable CPM by using the explicit form of $\tau^{(2)}$-matrices and Onsager algebra generators. Furthermore, by employing the theory of CPM, the nature of Onsager algebra representations and the full symmetry structure are revealed in the Bethe Ansatz approach using the explicit eigenvalues of superintegrable CPM. Note that in this paper, for simplicity, we have confined our study of  symmetry in superintegrable CPM only in the case of periodic boundary condition. We know that general skewed boundary conditions are required for the natural completion of the eigenvalue spectrum even in the Ising case. However, by results in \cite{B93}, the discussions of superintegrable CPM in this article can be carried over as well to skewed boundary conditions in general, hence the Onsager algebra symmetry would be valid for the complete spectrum of $\tau^{(j)}$-model.     
One may view the Onsager algebra symmetry of $\tau^{(2)}$-matrix in CPM as the counter part of the $sl_2$-loop algebra symmetry of the root-of-unity six-vertex model  found in \cite{DFM} within the scheme proposed in \cite{FM04} on the (conjectured) analogous theories between the eight vertex model and CPM. In this work, we have performed a rigorous study of the ${\cal P}$-polynomial (\req(ptv)), the pivotal ingredient in $\tau^{(j)}$-degeneracy, by explicit results of eigenvalues of CPM transfer matrix. In the context of $sl_2$-loop algebra symmetry of six-vertex model \cite{De05, FM01}, one has the Drinfeld polynomial, which plays the similar role of ${\cal P}$-polynomial in CPM, however is not fully understood till now. The exact results we have obtained in this work could be valuable to pave the way to solution of some conjectured symmetry problems in the root-of-unity six-vertex model. A process along this line is now under consideration. Moreover, we have made a detailed analysis on the comparison on fusion matrices and functional relations of the root-of-unity eight-vertex model and superintegrable CPM. As a consequence, we found  the relation (\req(8TT)) in eight-vertex model corresponding to $T\widehat{T}$-relations (\ref{TTS}) in CPM, which has played the vital role for the derivation of all functional relations in \cite{BBP}. 
However the quest for rigorous justification and physical interpretation of (\req(8TT)) has been left a challenge for the root-of-unity eight-vertex model as a completely parallel theory to CPM.  The use for deeper understanding of the $Q$-operators would be expected. Nevertheless, the striking similarity of these two theories would suggest the universal role of CPM in regard to the $"Q"$-operator in Baxter's TQ-relation, which appears also in other solvable lattice models. We made a attempt in Sec. 2 to explore the common features of functional equations for other plausible, but hypothetical, lattice models, based on the structure of superintegrable CPM listed in Sec. 1. The principle could be very sketchy, and in part just based on speculation; however, we hope it might help to give a feeling of how the current research on symmetry of $\tau^{(j)}$-models is oriented.

\section*{Acknowledgements}
The author is pleased to thank Professor Yum-Tong Siu for hospitality in April, 2005 at  Mathematical Department of Harvard University where part of this work was carried out. This work is supported in part by National Science Council of Taiwan under Grant No NSC 93-2115-M-001-003.

\end{document}